\documentclass[aps,pra,onecolumn,showpacs,superscriptaddress,floatfix,nofootinbib,10pt]{revtex4-1}

\usepackage{geometry}
\usepackage{makeidx}         
\usepackage{graphicx}        
\usepackage{multirow}
\usepackage[bottom]{footmisc}

\usepackage{amsmath}
\usepackage{epsfig}
\usepackage{helvet}
\usepackage{amssymb}

\newcommand{\ket}[1]{\mbox{$| #1 \rangle$}}
\newcommand{\bra}[1]{\mbox{$\langle #1 |$}}
\def\tr{ \mbox{tr}}

\begin{document}

\title{Quantum Criticality with the Multi-scale Entanglement Renormalization Ansatz}

\author{G. Evenbly}
\affiliation{The University of Queensland, Brisbane, Queensland 4072, Australia}
\email{glen.evenbly@gmail.com}

\author{G. Vidal}
\affiliation{Perimeter Institute for Theoretical Physics, Waterloo, Ontario N2L 2Y5, Canada}  \email{gvidal@perimeterinstitute.ca}

\begin{abstract}
The goal of this manuscript is to provide an introduction to the multi-scale entanglement renormalization ansatz (MERA) and its application to the study of quantum critical systems. Only systems in one spatial dimension are considered. The MERA, in its scale-invariant form, is seen to offer direct numerical access to the scale-invariant operators of a critical theory. As a result, given a critical Hamiltonian on the lattice, the scale-invariant MERA can be used to characterize the underlying conformal field theory. The performance of the MERA is benchmarked for several critical quantum spin chains, namely Ising, Potts, XX and (modified) Heisenberg models, and an insightful comparison with results obtained using a matrix product state is made. The extraction of accurate conformal data, such as scaling dimensions and operator product expansion coefficients of both local and non-local primary fields, is also illustrated. 
\end{abstract}

\pacs{05.30.-d, 02.70.-c, 03.67.Mn, 05.50.+q}

\maketitle
\tableofcontents



\section{Introduction} \label{sect:Intro}
Tensor network states are powerful variational ans\"atze that can be used to characterize the low-energy properties of quantum many-body systems on a lattice. The premise of tensor network approaches is to parameterize a many-body wave-function by using a collection of tensors connected into a network. The number of parameters required to specify these tensors is much smaller than the exponentially large dimension of the system's Hilbert space, in such a way that very large (and even infinite) lattices can be considered.

Tensor network states can be broadly classified into two sets according to the geometry of the underlying networks \cite{Evenbly11}. In the first set, the network reproduces the \emph{physical geometry} of the system, as specified by the pattern of interactions in the Hamiltonian. For instance, the matrix product state (MPS) \cite{Fannes92,Ostlund95,Rommer97}, an ansatz for $D=1$ dimensional systems, consists of a collection of tensors connected into a chain; similarly, its generalization for lattices in $D>1$ dimensions, known as projected entangled pair states (PEPS) \cite{Verstraete04,Murg07,Jordan08}, consists of a collection of tensors connected according to a $D$ dimensional lattice. In contrast, a second class of tensor network states aim at reproducing the \emph{holographic geometry} of the system. The latter spans an additional dimension used to parameterize the different length scales (or, equivalently, energy scales) relevant to the description of the many-body wave-function. Thus the multi-scale entanglement renormalization ansatz (MERA) \cite{Vidal07, Vidal08, Evenbly10,Cincio08,Evenbly10b,Aguado08,Giovannetti08, Evenbly09, Pfeifer09,Evenbly09b, Koenig09,Montangero09,Vidal10, Evenbly10c, Evenbly10d, Cincio10} for a lattice system in $D$ dimensions consists of a network in $D+1$ dimensions.

The simplest and most widely employed tensor network state is the MPS. The MPS underlies the remarkable success of the density matrix renormalization group (DMRG) algorithm \cite{White92, White93, Schollwoeck05, Schollwoeck11}, which for almost two decades has dominated the numerical study of $D=1$ dimensional quantum systems, providing very accurate results for low energy properties. DMRG has not only become ubiquitous in condensed matter physics but has also found application in other fields involving quantum many-body systems, such as quantum chemistry \cite{Chan08}. Further algorithms based upon the MPS have also been developed, such as the time evolving block decimation (TEBD) \cite{Vidal03, Vidal04} algorithm, later reformulated as time-dependent DMRG \cite{Daley04, White04}, which allows the simulation of certain low-energy \emph{dynamics} for relatively long times.

In this manuscript we discuss the application of the MERA to study critical systems in $D=1$ dimensions  (although most of the present formalism is directly applicable to $D>1$ dimensions). Given the success of MPS-based methods such as DMRG and TEBD for quantum systems in $D=1$ dimensions, it is natural to ask whether an alternative approach is actually needed. A clear answer to this question is obtained by discussing the short-comings of the MPS representation for critical systems as well as by exploring the benefits of including scale invariance directly into a tensor network state, something that is possible with the MERA but not the MPS.

Critical systems typically lack a characteristic length scale and are thus invariant under changes of scale. One manifestation of scale invariance in critical systems is that correlators decay polynomially, in sharp contrast with gapped systems, where they decay exponentially with a characteristic correlation length. It is well-known, however, that an MPS with a finite bond dimension $\chi$ (where $\chi$ indicates the size of the tensors) can never properly capture the scale invariance of a critical state. Indeed, a duly optimized MPS possesses an intrinsic finite correlation length $\zeta \approx \chi^\kappa$ \cite{Tagliacozzo08,Pollmann09}, where $\kappa$ is a constant that depends on the universality class of the phase transition under consideration, such that correlators decay exponentially at length scales larger than $\zeta$. Thus, while the MPS can accurately approximate short-range properties of a critical ground state, it necessarily fails to capture its properties at asymptotically large distances. [In practice, however, the cost of MPS-based approaches scales only as $O(\chi^3)$ with the bond dimension $\chi$. This means that one can use a very large value of $\chi$, which often allows the critical behavior of a system to be accurately captured up to some very large length scale $\zeta$.]

On the other hand, the MERA can explicitly capture the scale invariance of critical systems \cite{Vidal07, Evenbly10, Evenbly10b, Evenbly09, Pfeifer09, Montangero09, Evenbly09b, Evenbly10d}, a feature that has significant advantages, both conceptual and practical. Tensors in the MERA are organized in layers, where each layer corresponds to a different length (or energy) scale. In an infinite system, scale invariance is then easily enforced by choosing all layers of tensors to be identical. The resulting ansatz is referred to as the \emph{scale-invariant} MERA. Certain structural properties of the scale-invariant MERA, such as the polynomial decay of correlators and the logarithmic growth of block entanglement entropy \cite{Evenbly11,Vidal08}, already hint at its suitability to represent critical systems. 

In addition, this ansatz offers direct access to the scaling operators of a critical theory, namely those operators that transform into themselves under scale transformations. As a scale-invariant/covariant object, a scaling operator must act non-trivially on a region of the system that has no characteristic length scale. In a $(1+1)$-dimensional conformal field theory (CFT) \cite{Cardy96, Francesco97, Henkel99} (corresponding to the continuum limit of a critical quantum system in $D=1$ spatial dimensions), the support of a scaling operator can therefore only be one of three possibilities: (i) an infinite line, (ii) a single point,  or (iii) a semi-infinite line. The first type of support corresponds to a global internal symmetry of the CFT's Hamiltonian. The second type of support is seen to correspond to local scaling operators, associated to local excitations. Finally, the third type corresponds to non-local (or semi-local) scaling operators, associated e.g. to domain wall excitations. Going back to the lattice, scaling operators are distorted by the presence of a finite lattice spacing, but they can still be directly extracted from the scale-invariant MERA. Thus, on the lattice, (i) a global internal symmetry is implemented by an infinite string of identical single-site operators; (ii) local scaling operators are now supported on a small number of sites (the specific number depends on the MERA scheme); and (iii) non-local operators mix elements of the two previous objects: they consist of a semi-infinite string of identical single-site operators (the same ones that implement an internal symmetry) completed with some local operator at the end of the string. Importantly, the scaling dimensions and fusion rules of the scaling operators on the lattice are the same as in the continuum. As a result, a relatively simple and inexpensive calculation with the scale-invariant MERA can be used to obtain remarkably accurate estimates of the conformal data characterizing the underlying CFT.

The rest of the manuscript is organized in sections as follows. Sect. \ref{sect:ERandMERA} introduces the key aspects of entanglement renormalization, which is the renormalization group transformation for lattice models on which the MERA is based. We describe how local operators transform under the coarse-graining transformation and discuss basic aspects of the MERA, including the evaluation of the expectation value of local observables, and briefly compare different MERA implementations in $D=1$ dimensions.
Sect. \ref{sect:SymMERA} addresses the role of spatial and internal symmetries in tensor network states, and compares how different symmetries can be enforced in MPS/PEPS and MERA. While translation invariance is naturally implemented in an MPS and PEPS, it can only be approximately enforced on the MERA. Sect. \ref{sect:ScaleMERA} specializes on the implementation of scale invariance in the MERA and discusses how the scaling operators of a critical theory can be extracted from it.
In Sect. \ref{sect:MERABench} we demonstrate the performance of the scale-invariant MERA for a number of critical quantum spin chains, namely Ising, Potts, quantum XX and Heisenberg models. Specifically, Sect. \ref{sect:MERAvsMPS} compares ground state energy and two-point correlators obtained with MERA and MPS. Interestingly, MPS and MERA approaches seem to complement each other. For a given computational cost, the MPS is seen to provide more accurate estimates of the expectation value of a local observable, such as the ground state energy. However, the MERA is seen to provide a better characterization of long-range properties, such as two-point correlators at long distances. The advantages of the scale-invariant MERA are then further illustrated in Sect. \ref{sect:CFTBench} by extracting, in the concrete context of the quantum Ising model, the conformal data (including scaling dimensions, fusion rules for scaling operators and central charge) of the underlying CFT.

\begin{figure}[!htbp]
\begin{center}
\includegraphics[width=11cm]{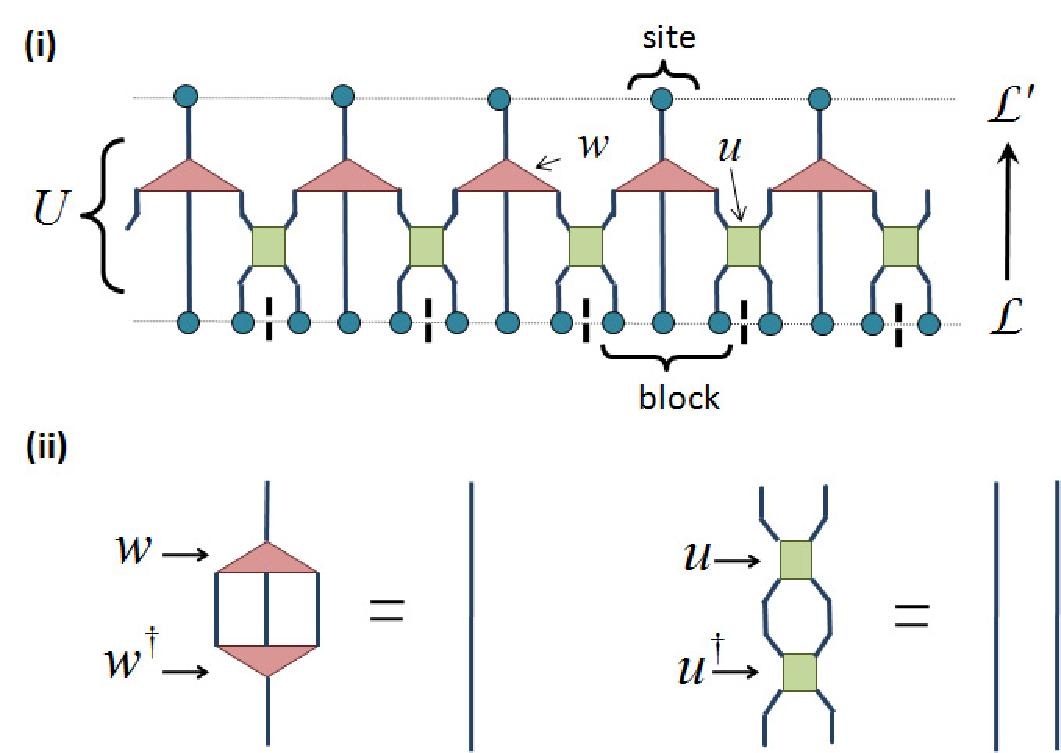}
\caption{(i) The coarse-graining transformation $U$, a specific implementation of entanglement renormalization, is comprised of isometries $w$ and disentanglers $u$ and maps \emph{blocks} of three sites from the initial lattice $\mathcal L$ into a \emph{site} of the coarser lattice $\mathcal L '$. (ii) The tensors $w$ and $u$ are constrained to be isometric, see also Eq. \ref{eq:sRe3}.}
\label{fig:ERandMERA}
\end{center}
\end{figure}

\section{Entanglement Renormalization and the MERA} \label{sect:ERandMERA}

In this section we first recall a few basic aspects of entanglement renormalization (ER), the coarse-graining transformation upon which the MERA is based. We then introduce the MERA and review a few of its features.

\subsection{Foundations of Entanglement Renormalization} \label{sect:ERfoundation}
For concreteness, we mostly consider a specific implementation of ER that produces the so-called ternary MERA (where three sites are coarse-grained into one effective site). In Sect. \ref{sect:ChoiceScheme} we also discuss other MERA schemes. 

Let $\mathcal{L}$ denote a $D=1$ dimensional lattice made of $N$ sites, where each site is described by a Hilbert space $\mathbb{V}$ of finite dimension $d$, so that the vector space of the lattice is $\mathbb{V}_{ \mathcal{L}} \cong \mathbb{V}^{\otimes N}$. We consider a coarse-graining transformation $U$ that maps lattice $\mathcal L$ to a coarser lattice $\mathcal L'$ 
\begin{equation}
U^{\dagger}:\mathbb V_{\mathcal{L}}  \mapsto \mathbb V_{\mathcal L'}. \label{eq:sRe1}
\end{equation}
where $\mathcal L'$ is made of $N/3$ sites, each with a vector space $\mathbb{V}'$ of dimension $\chi$, so that $\mathbb{V}_{ \mathcal{L}'} \cong \mathbb{V}'^{\otimes N/3}$, and where transformation $U$ decomposes into local transformations, known as disentanglers $u$ and isometries $w$,
\begin{equation}
u^{\dagger}:\mathbb V^{ \otimes 2}  \mapsto \mathbb V^{ \otimes 2} ,\; \; \; w^{\dagger}:\mathbb V^{ \otimes 3}  \mapsto \mathbb V' \label{eq:sRe2},
\end{equation}
according to Fig. \ref{fig:ERandMERA}(i). More specifically, if we partition the initial lattice $\mathcal L$ into \emph{blocks} of three sites, then the disentanglers $u$ are first applied across the boundaries of neighboring blocks, followed by the isometries $w$, which map each block of three sites into a single effective site of the coarser lattice $\mathcal L'$. Disentanglers and isometries are required to satisfy isometric constraints, namely
\begin{equation}
u^\dag  u = \mathbb I ^{ \otimes 2} ,\; \; \; w^\dag  w = \mathbb I' \label{eq:sRe3},
\end{equation}
where $\mathbb I$ and $\mathbb I'$ are the identity operator in $\mathbb V$ and $\mathbb V'$, respectively, see Fig. \ref{fig:ERandMERA}(ii). Note that, by construction, the disentangler $u$ is also unitary, that is $u u^\dag = \mathbb I^{ \otimes 2}$. The dimension $\chi$ of the Hilbert space $\mathbb{V}'$ can be chosen to be different than $d$, provided that $\chi \le d^3$ (as demanded by the above isometric constraint on $w$). In general, choosing a larger dimension $\chi$, i.e. retaining a larger effective Hilbert space $\mathbb{V}'$ for each coarse-grained site, yields a more accurate RG transformation, one that better preserves the low energy properties of the system.  

\begin{figure}[!htb]
\begin{center}
\includegraphics[width=12cm]{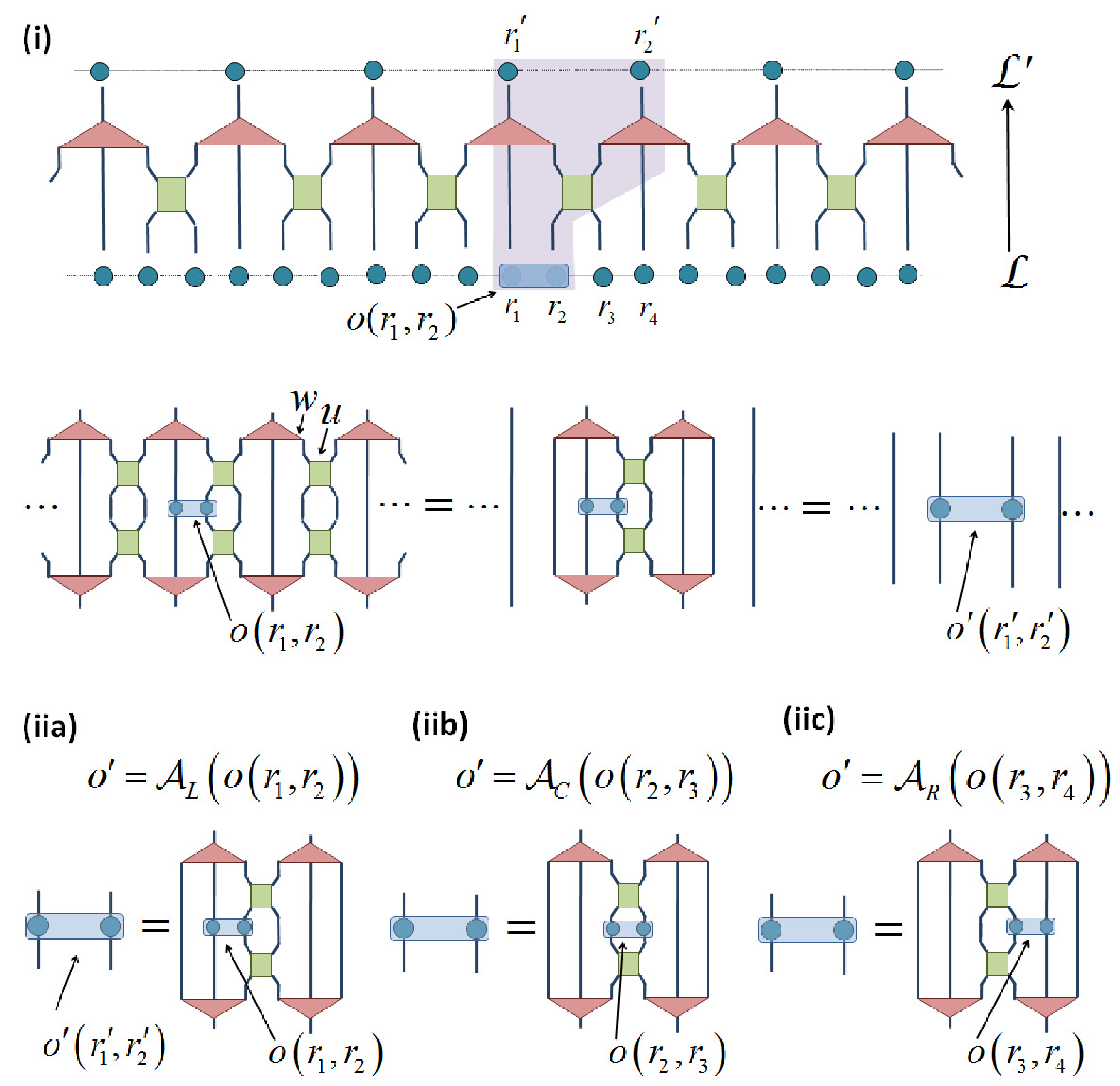}
\caption{(i) Under coarse-graining with entanglement renormalization, an operator $o(r_1, r_2)$ supported on two sites of lattice $\mathcal{L}$ is transformed into a new operator $o'(r_1 ', r_2 ')$ supported on two sites of the coarser lattice $\mathcal{L '}$, see Eq. \ref{eq:sRe4}. The coarse-graining of local operators can be implemented directly via the (iia) left, (iib) center and (iic) right ascending superoperators, denoted $\mathcal{A}_L, \mathcal{A}_C$ and $\mathcal{A}_R $ respectively. Notice that the coarse-graining of $o(r_1, r_2)$ in (i) corresponds to application of the left ascending superoperator $\mathcal{A}_L$.}
\label{fig:CGlocal}
\end{center}
\end{figure}

An important property of the coarse-graining transformation $U$ is that it preserves \emph{locality}.
Let $o(r_1, r_2)$ be a local operator defined on two contiguous sites $(r_1, r_2)$ of lattice $\mathcal{L}$. Under coarse-graining, the operator $o(r_1, r_2)$ becomes
\begin{equation}
U^\dag  o(r_1, r_2) U = \cdots \otimes \mathbb I' \otimes o'(r_1 ' , r_2 ')  \otimes \mathbb I' \otimes \cdots , \label{eq:sRe4}
\end{equation}
where the only non-trivial part of $U^\dag  o(r_1, r_2) U$ is a new operator $o'(r_1',r_2')$ supported on two contiguous sites $(r_1 ' , r_2 ')$ of lattice $\mathcal{L}'$. Notice that operator $o'(r_1',r_2')$ remains local (i.e., it is supported on two sites) thanks to both the specific decomposition of $U$ into disentanglers and isometries, and the isometric constraints of these tensors, Eq. \ref{eq:sRe3}, which ensure that most of the tensors of $U$ in $U^\dag  o(r_1, r_2) U$ annihilate to identity with their conjugates in $U^\dag$, as shown Fig. \ref{fig:CGlocal}(i). In view of this fact, it is most convenient to introduce left, center and right \textit{ascending superoperators} $\left\{ \mathcal{A}_L, \mathcal{A}_C, \mathcal{A}_R \right\}$, as shown Fig. \ref{fig:CGlocal}(ii), which directly produce the two-site coarse-grained operator $o'$ from the two-site operator $o$, as given by one of the following,  
\begin{align}
 o'\left({r'_1 ,r'_2 } \right) &= \mathcal{A}_L \left( {o\left( {r_1 ,r_2 } \right)} \right), \nonumber\\ 
 o'\left( {r'_1 ,r'_2 } \right) &= \mathcal{A}_C \left( {o\left( {r_2 ,r_3 } \right)}\right), \nonumber\\ 
 o'\left( {r'_1 ,r'_2 } \right) &= \mathcal{A}_R \left( {o\left( {r_3 ,r_4 } \right)}\right), \label{eq:sRe5} 
\end{align}
where the specific choice of ascending superoperator to be used depends on the location of the operator $o$ on the lattice $\mathcal{L}$. 

We may now concatenate the coarse-graining transformation a number $T$ of times to obtain a \emph{sequence} of coarser lattices,
\begin{equation}
\mathcal L^{[0]} \stackrel{U^{[0]}}{\longmapsto} \mathcal L^{[1]} \stackrel{U^{[1]}}{\longmapsto}  \cdots \stackrel{U^{[T-1]}}{\longmapsto} \mathcal L^{[T]}, \label{eq:sRe6}
\end{equation}
where we use superscripts in square brackets to denote the level of coarse-graining, with the initial lattice $\mathcal L^{[0]} \equiv \mathcal L$. Then, for any local (i.e., two-site) operator $o^{[0]}\equiv o$ defined on $\mathcal L^{[0]}$, the transformations $\{U^{[\tau]}\}$ generate a sequence of local coarse-grained operators $\{o^{[\tau]}\}$, defined on lattices $\{\mathcal L^{[\tau]}\}$, 
\begin{equation}
o^{[0]} \stackrel{\mathcal A^{[0]}}{\longmapsto} o^{[1]} \stackrel{\mathcal A^{[1]}}{\longmapsto}  \cdots \stackrel{\mathcal A^{[T-1]}}{\longmapsto} o^{[T]}, \label{eq:sRe7}
\end{equation}
through application of the appropriate ascending superoperator $\mathcal A$, as per Eq. \ref{eq:sRe5}.

\begin{figure}[!tbhp]
\begin{center}
\includegraphics[width=12cm]{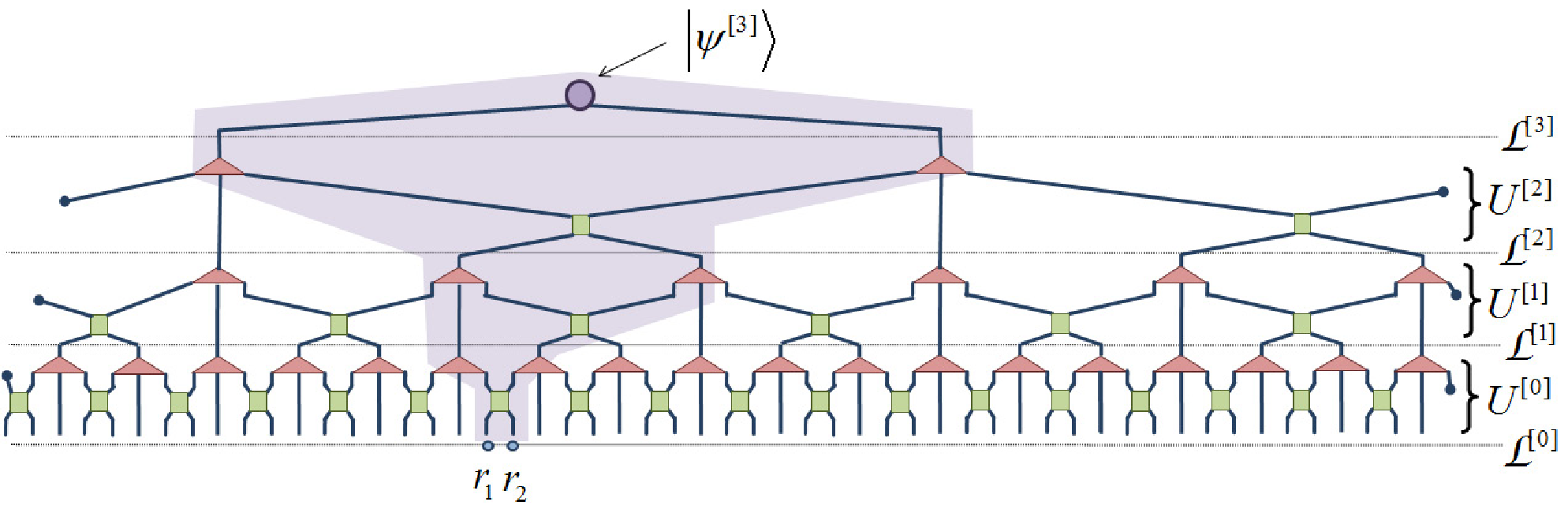}
\caption{The ternary MERA for a lattice $\mathcal L^{[0]}$ of $N=54$ sites. Each layer $U^{[\tau]}$ of the MERA can be interpreted as a coarse-graining transformation between an initial lattice $\mathcal L ^{[\tau]}$ and a coarser lattice $\mathcal L ^{[\tau+1]}$. The \emph{past causal cone} of two sites $(r_1,r_2)$ in lattice $\mathcal L^{[0]}$ is shaded.}
\label{fig:TernMERA}
\end{center}
\end{figure}

\subsection{Foundations of the MERA} \label{sect:MERAfoundation}

We have just seen that the ER transformation $U$ can be used to coarse-grain local operators, producing a renormalization group (RG) flow for local operators, Eq. \ref{eq:sRe7}. As a linear (isometric) map from $\mathbb{V}_{\mathcal L}$ to $\mathbb{V}_{\mathcal L '}$, $U$ can of course also be used to coarse-grain quantum states. More important for us, however, is to consider an \emph{inverse} RG flow of states. Let us assume that we have the sequence of ER transformations $\left\{ U^{[0]}, U^{[1]}, \ldots ,U^{[T-1]} \right\}$ which act on an initial lattice $\mathcal L^{[0]}$ of $N$ sites to eventually produce coarse-grained lattice $\mathcal L^{[T]}$. Then for a quantum state $|\psi^{[T]}\rangle$ defined on lattice $\mathcal L^{[T]}$, the transformation $U^{[T-1]}$ can be used to obtain a new state $|\psi^{[T-1]}\rangle$,
\begin{equation}
\left| {\psi^{[T-1]}} \right\rangle \equiv  U^{[T-1]}  \left| \psi^{[T]}  \right\rangle, \label{eq:sRe8}
\end{equation}
defined on the finer lattice $\mathcal L^{[T-1]}$. Through iteration of Eq. \ref{eq:sRe8}, one can further obtain increasingly fine-grained states, eventually reaching a state $|\psi^{[0]}\rangle$ defined on $\mathcal L^{[0]}$,
\begin{equation}
\left| {\psi ^{[0]} } \right\rangle  = U^{[0]} U^{[1]}  \cdots U^{[T - 1]} \left| {\psi ^{[T]} } \right\rangle. \label{eq:sRe9}
\end{equation}
Let us assume that the number of levels $T$ is chosen $T \approx \log_3 (N)$, such that the maximally coarse-grained lattice $\mathcal L^{[T]}$ contains a small number of sites and hence the state $\left| {\psi ^{[T]} } \right\rangle$ can also be described with a small number of parameters. Then the multi-scale entanglement renormalization ansatz (MERA) is the class of states $\left| {\psi ^{[0]} } \right\rangle$ that can be represented as Eq. \ref{eq:sRe9}, for some choice of $\left\{ U^{[0]}, U^{[1]}, \ldots ,U^{[T-1]} \right\}$ and $|\psi^{[T]}\rangle$. For instance, Fig. \ref{fig:TernMERA} depicts the MERA, organized into $T=3$ layers, for a state $\left| {\psi ^{[0]} } \right\rangle$ on a lattice of $N=54$ sites,
\begin{equation}
\left| {\psi ^{[0]} } \right\rangle  = U^{[0]} U^{[1]} U^{[2]} \left| {\psi ^{[3]} } \right\rangle. \label{eq:sRe10}
\end{equation}

Let us now count variational parameters. We will assume for simplicity that for any value of $\tau = 0,1, \cdots,T-1$, the dimension of the vector space $\mathbb{V}^{[\tau]}$ is $\chi$. Recall that the transformations $\{U^{[\tau]}\}$ themselves are comprised of local tensors, the disentanglers $u$ and isometries $w$, each specified by $\chi^4$ parameters. Since in an $N$-site lattice there are $O(N)$ disentanglers $u$ and isometries $w$ (distributed in $O(\log (N))$ layers), the MERA depends on $O(N\chi^4)$ parameters. In Sect. \ref{sect:SymMERA} we will see that, through incorporation of spatial symmetries into the ansatz, this number can be reduced to $O(\chi^4)$, which is independent of $N$ and allows for the study of infinite systems. 

\begin{figure}[!tbph]
\begin{center}
\includegraphics[width=12cm]{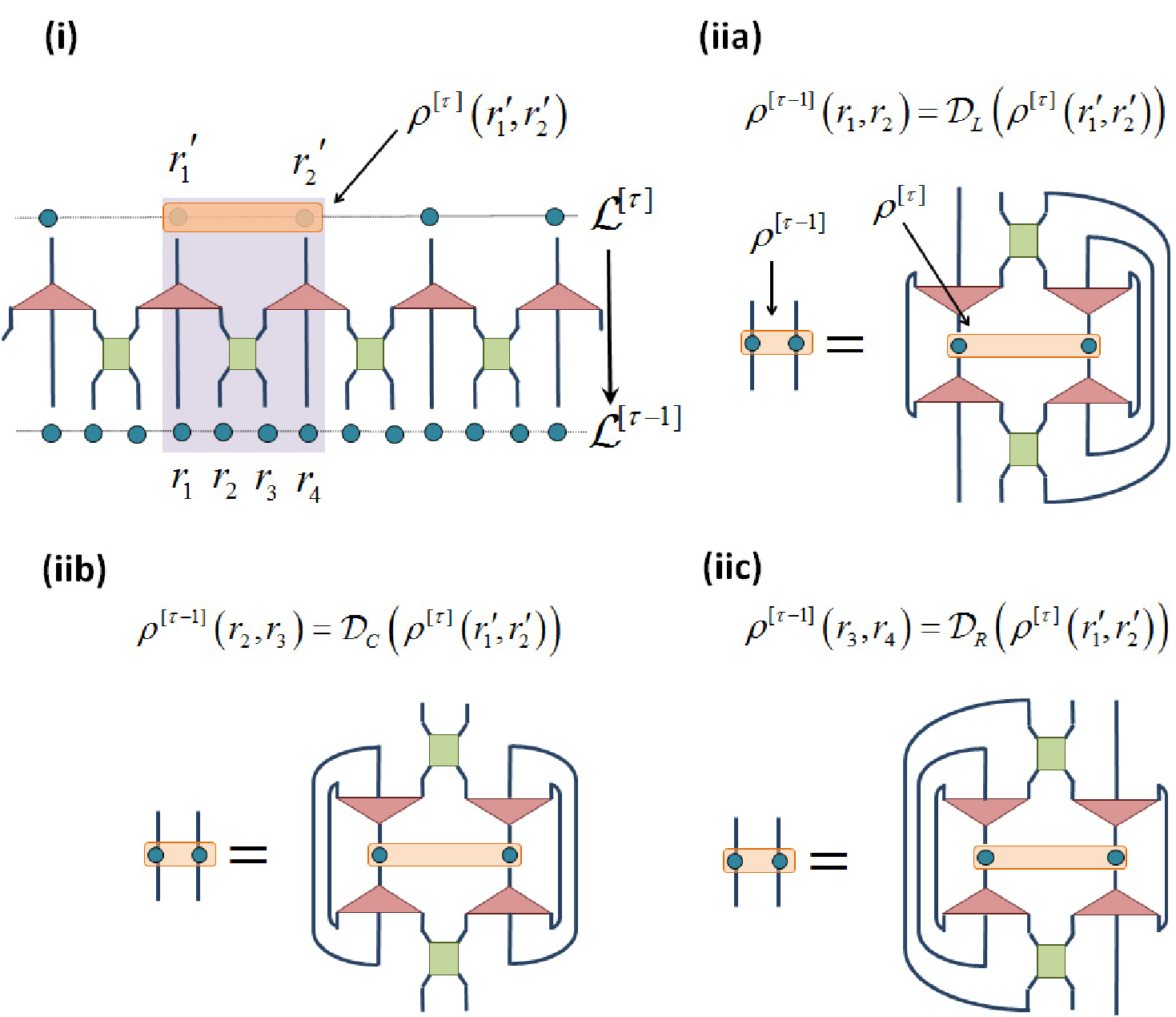}
\caption{(i) The causal cone (shaded) of four sites $(r_1, r_2, r_3, r_4)$ in lattice $\mathcal L^{[\tau-1]}$ involves two sites $(r_1 ', r_2 ')$ in lattice $\mathcal L^{[\tau]}$. Starting from the reduced density matrix $\rho^{[\tau]} (r_1 ', r_2 ')$ on lattice $\mathcal L^{[\tau]}$, the reduced density matrix on any pair of contiguous from $(r_1, r_2, r_3, r_4)$ can be obtained using the (iia) left, (iib) center and (iic) right descending superoperators, denoted $\mathcal{D}_L, \mathcal{D}_C, \mathcal{D}_R $ respectively.}
\label{fig:DensityLower}
\end{center}
\end{figure}

We have therefore established that the MERA can be specified with a number of parameters that is much smaller than the dimension of the Hilbert space $\mathbb{V}_{\mathcal L}$, which grows exponentially in the number $N$ of sites. However, for this ansatz to be useful, we also need to be able to efficiently extract information about $|\psi^{[0]}\rangle$ in Eq. \ref{eq:sRe9}. For any local operator $o^{[0]}(r_1,r_2)$, and due to the very peculiar causal structure of its tensor network, it is actually possible to efficiently compute the expectation value
\begin{equation}
\left\langle o^{[0]}(r_1,r_2) \right\rangle  \equiv \left\langle \psi^{[0]} \right| o^{[0]}(r_1,r_1)  \left| \psi^{[0]} \right\rangle
\end{equation}
from the MERA. Let us define the \emph{past causal cone} of a site in lattice $\mathcal L^{[\tau]}$ as the set of tensors and indices that can affect the state on that site. By construction, in a MERA the causal cone of any site of $\mathcal L^{[\tau]}$ is seen to involve just a constant (that is, independent of $N$) number of sites of any other lattice $\mathcal{L}^{[\tau']}$ for $\tau'>\tau$, a property that we refer to by saying that the past causal cone has bounded `width'. Fig. \ref{fig:TernMERA} displays the past causal cone of two sites $(r_1 , r_2)$ in a ternary MERA, which only involves two sites of every lattice $\mathcal L^{[\tau]}$. This property allows for the efficient computation of local reduced density matrix $\rho^{[0]}(r_1,r_2)$, from which the expectation value 
\begin{equation}
\left\langle o^{[0]}(r_1,r_1) \right\rangle  = {\rm{tr}}\left( {o^{[0]}(r_1,r_1) \rho^{[0]}(r_1,r_2) } \right) \label{eq:sRe13}
\end{equation}
can be obtained. The computation of local reduced density matrices is simplified through the introduction of left, center and right descending superoperators, $\left\{ \mathcal{D}_L, \mathcal{D}_C, \mathcal{D}_R \right\}$, which are the adjoints of the ascending superoperators of Eq. \ref{eq:sRe5}. Let us assume that we have the density matrix $\rho^{[\tau]} (r_1 ', r_2 ')$ describing the state on two contiguous sites $(r_1 ', r_2 ')$ of lattice $\mathcal L^{[\tau]}$. Then, as shown Fig. \ref{fig:DensityLower}, the descending superoperators may be used to compute the two-site reduced density matrix $\rho^{[\tau-1]}$ on certain sites of the lattice $\mathcal L^{[\tau-1]}$,
\begin{align}
\rho^{[\tau-1]}(r_1, r_2 ) &= \mathcal D_L \left(  \rho^{[\tau]} (r_1 ', r_2 ') \right),\nonumber \\ 
\rho^{[\tau-1]}(r_2, r_3 ) &= \mathcal D_C \left(  \rho^{[\tau]} (r_1 ', r_2 ') \right),\nonumber \\
\rho^{[\tau-1]}(r_3, r_4 ) &= \mathcal D_R \left(  \rho^{[\tau]} (r_1 ', r_2 ') \right). \label{eq:sRe11}
\end{align}
Thus, through repeated use of the appropriate descending superoperator in Eq. \ref{eq:sRe11}, we can compute the reduced density matrix $\rho^{[0]}(r_1, r_2)$ of any two contiguous sites $(r_1, r_2)$ of the original lattice $\mathcal L^{[0]}$ by `lowering' the density matrix through the appropriate causal cone. For instance, the reduced density matrix $\rho^{[0]} (r_1 , r_2)$ on the two sites $(r_1 , r_2)$ of lattice $\mathcal L^{[0]}$ shown in Fig. \ref{fig:TernMERA} can be computed as
\begin{equation}
\rho^{[0]} (r_1, r_2)  = {\cal D}_C \left( {{\cal D}_L \left( {{\cal D}_C \left( {\left| {\psi ^{(3)} } \right\rangle \left\langle {\psi ^{(3)} } \right|} \right)} \right)} \right). \label{eq:sRe12}
\end{equation}
The cost of calculating a local reduced density matrix $\rho^{[0]}$, as in the example of Eq. \ref{eq:sRe12}, is proportional to the number $T\approx \log (N)$ of layers in the MERA, hence scales with system size $N$ as $O(\log (N))$.  Two point correlators can also be evaluated using similar manipulations, see Ref. \onlinecite{Evenbly09}.

\begin{figure}[!p]
\begin{center}
\includegraphics[width=12cm]{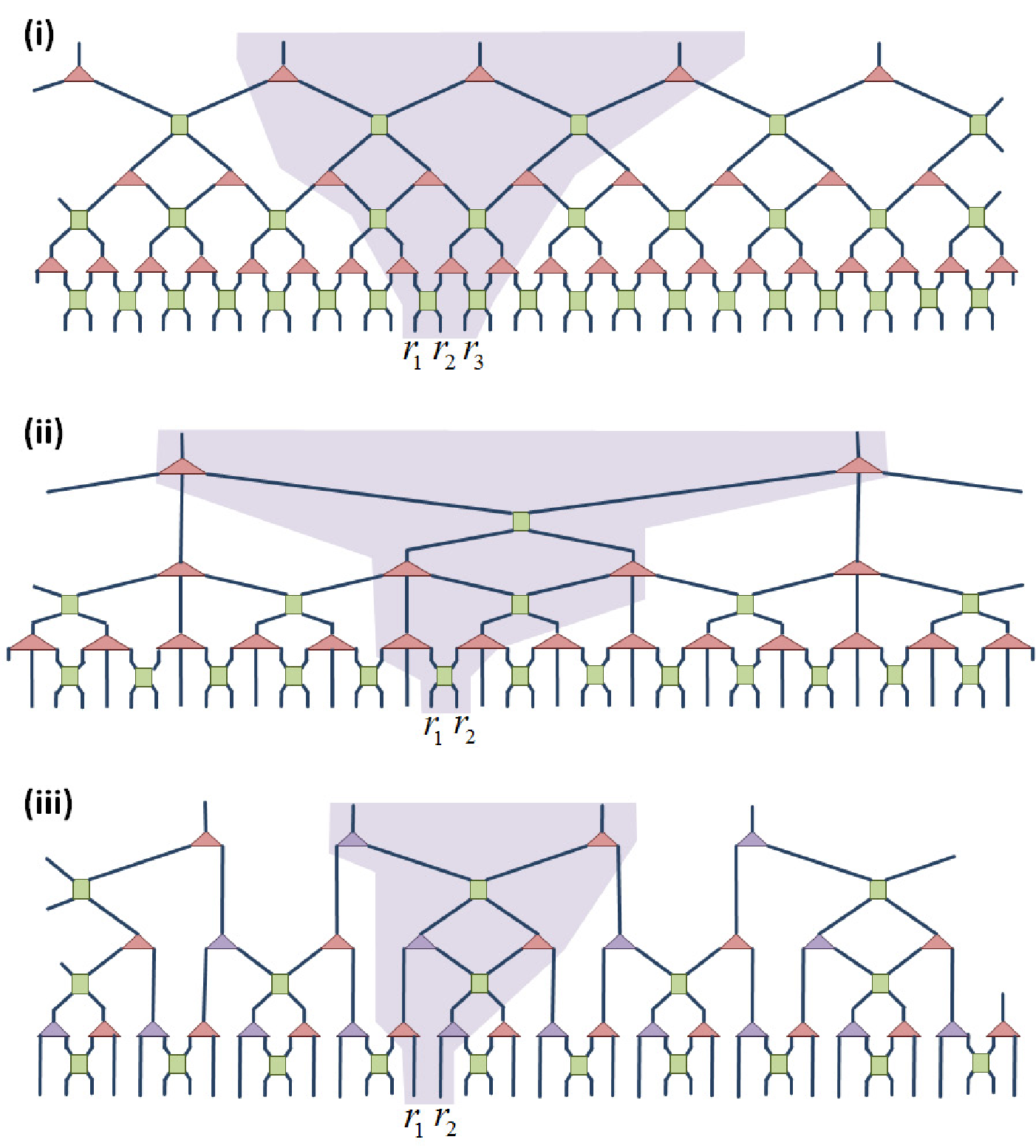}
\caption{Three different MERA schemes for a $D=1$ dimensional lattice. An example of a causal cone is shaded for each scheme. (i) The binary MERA scheme, based upon a 2-to-1 coarse-graining step, has a causal width of three sites and a cost of contraction that scales with the bond dimension $\chi$ as $O(\chi^9)$. (ii) The ternary MERA scheme, based upon a 3-to-1 coarse-graining step, has a causal width of two sites and a cost of contraction that scales as $O(\chi^8)$. (iii) The modified binary MERA scheme, equivalent to the binary MERA scheme with every second disentangler removed, has a causal width of two sites and a cost of contraction that scales as $O(\chi^7)$.}
\label{fig:SchemeMERA}
\end{center}
\end{figure}

\subsection{Choice of MERA Scheme} \label{sect:ChoiceScheme}

There are many possible ways of implementing the MERA in $D=1$ dimensions, of which the ternary MERA described in Sect. \ref{sect:MERAfoundation} is just one example. Fig. \ref{fig:SchemeMERA} displays a ternary MERA together with two other possible implementations: the binary MERA scheme (in terms of which the first ER proposals \cite{Vidal07,Vidal08} were formulated) and a modified binary MERA scheme (with half the amount of disentanglers as the previous binary scheme). While all MERA schemes function similarly on a conceptual level, the computational efficiency may differ between schemes. For instance, the cost of evaluating the expectation value of local observables as described Sect. \ref{sect:MERAfoundation} scales, in terms of the bond dimension $\chi$, as $O(\chi^9)$ for the binary MERA, as $O(\chi^8)$ for the ternary MERA and as $O(\chi^7)$ for the modified binary MERA. On the other hand, the binary MERA scheme has more disentangling power than either the ternary or modified binary schemes and, for any given $\chi$, produces a more accurate representation of ground states. It is therefore not obvious which MERA implementation will give the best numeric results for a fixed computational budget. However, a direct comparison of performance, see Sect. \ref{sect:SchemeCompare}, shows that the modified binary scheme of Fig. \ref{fig:SchemeMERA}(iii) is the most efficient scheme. Consequently, this scheme is used for the obtaining the numeric results presented in Sect. \ref{sect:MERABench}. However, for the sake of simplicity, we shall continue to discuss theoretical aspects of MERA in terms of the ternary MERA.

\section{Symmetries in Tensor Network States} \label{sect:SymMERA}

Consider a many-body state that is invariant under some symmetry transformation. In approximating this state with a tensor network state, we would like to preserve the original symmetry as much as possible. In this section we examine the types of symmetries that can be enforced upon the MPS and PEPS, and upon the MERA. We also examine whether the presence of the symmetry can be exploited for computational gain. We begin by discussing spatial symmetries, followed by global internal symmetries. The results are summarized in Table \ref{tab:SymMPS}.

\begin{table}[!htbp]
\centering
\caption[]{Several symmetries that can be exactly enforced and/or whose presence can be exploited for computational gain with MPS/PEPS and MERA algorithms.}
\renewcommand{\arraystretch}{1.2}
\setlength\tabcolsep{5pt}
\begin{tabular}{ |c | c | c l |}
\multicolumn{4}{l}{\textbf{Symmetries with MPS/PEPS:}}\\
\hline
$~~~~~$                              & Enforceable      & \multicolumn{2}{l|}{Exploitable} \\
\hline
Translation invariance                 & Yes      & Yes,       & cost: $O(N)\rightarrow O(1)$\\
Scale invariance                       & Unlikely       & Unlikely        & $~~~~~$                       \\
Internal symmetries                  & \multirow{2}{*}{Yes} & \multirow{2}{*}{Yes}& \multirow{2}{*}{~~}\\
(e.g. $\mathbb Z_2, U(1), SU(2)$)    &      &                              &      \\
\hline\noalign{\bigskip}

\multicolumn{4}{l}{\textbf{Symmetries with MERA:}}\\
\hline
$~~~~~$                              & Enforceable      & \multicolumn{2}{l|}{Exploitable}  \\
\hline
Translation invariance                 & Unknown  & Yes,     & cost: $O(N)\rightarrow O(\log (N))$\\
Scale invariance                       & Yes      & Yes,     & cost: $O(\log (N))\rightarrow O(1)$\\
Internal symmetries                  & \multirow{2}{*}{Yes} & \multirow{2}{*}{Yes}& \multirow{2}{*}{~~}\\
(e.g. $\mathbb Z_2, U(1), SU(2)$)    &      &                              &      \\
\hline
\end{tabular}
\label{tab:SymMPS}       
\end{table}

\begin{figure}[!htbp]
\begin{center}
\includegraphics[width=12cm]{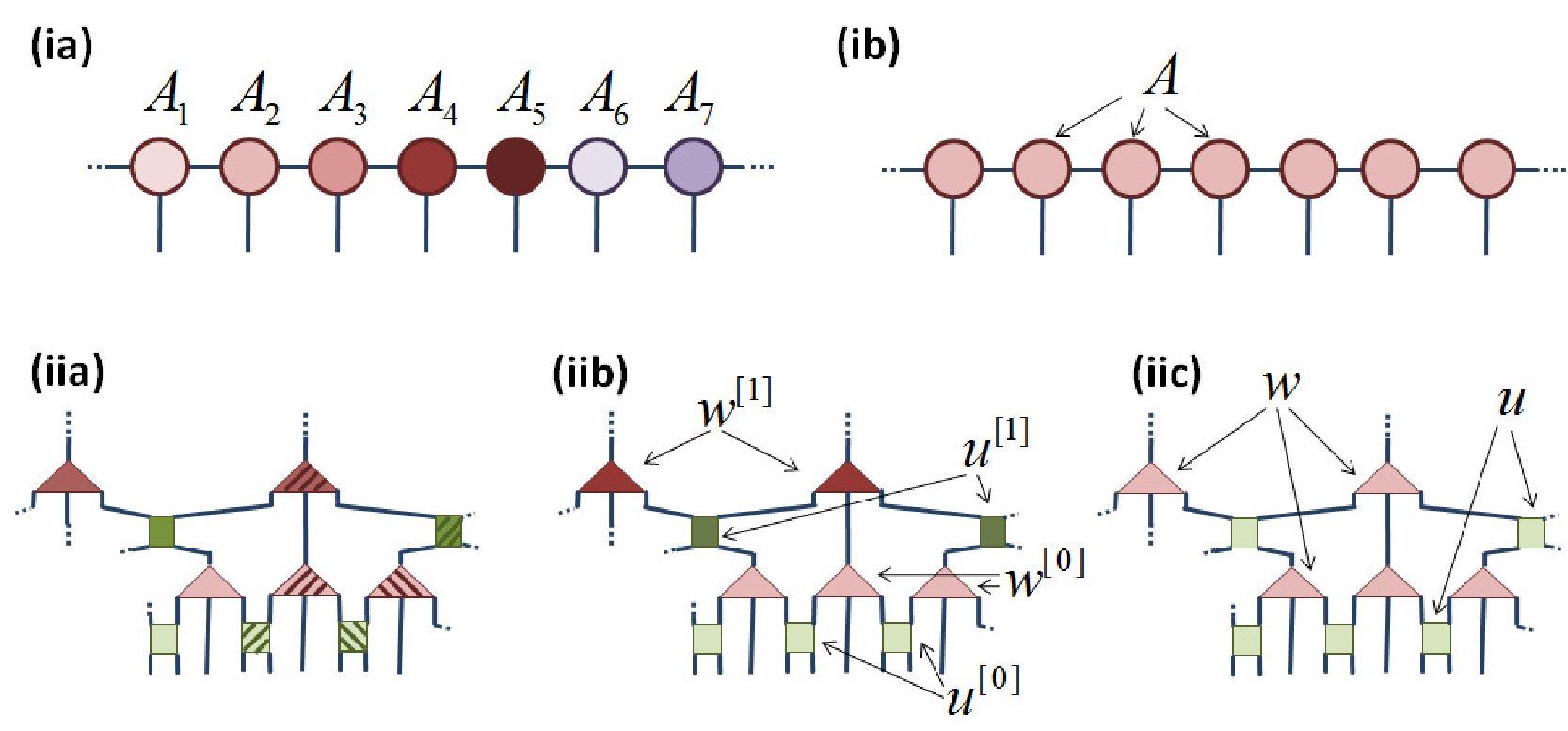}
\caption{(ia) An inhomogeneous MPS has an independent tensor $A_{\vec r}$ associated to each lattice site. (ib) All the tensors in a translation invariant MPS are chosen to be copies of the same tensor $A$. (iia) A generic MERA on an $N$ site lattice contains $O(N)$ different tensors. (iib) Translation invariance can be exploited by choosing the tensors in each layer $U^{[\tau]}$ of the MERA as copies of a single unique disentangler $u^{[\tau]}$ and isometry $w^{[\tau]}$. For an $N$ site lattice, this MERA contains $O(\log N)$ different tensors. (iic) Scale invariance can be incorporated into the MERA by further enforcing all layers to be identical, hence the entire MERA is described by a single $u$ and $w$.}
\label{fig:SpatialMERA}
\end{center}
\end{figure}

\subsection{Spatial Symmetries} \label{sect:SymSpatial}

For simplicity we discuss only two typical spatial symmetries: invariance under translations in homogeneous systems and invariance under changes of scale in e.g. critical systems. Let us first consider them in the context of the MPS for $D=1$ dimensions and PEPS for $D>1$ dimensions. 

In an inhomogeneous MPS/PEPS, one associates a different tensor $A_{\vec{r}}$ to each site $\vec{r}$ of the lattice $\mathcal{L}$. Hence, in a lattice made of $N$ sites, the total number of tensors in the tensor network is also $N$. In the presence of translation invariance (either in a finite system with periodic boundary conditions or in an infinite system), this symmetry can be incorporated into the MPS/PEPS by choosing all the tensors to be a copy of the same tensor $A$, i.e. $A_{\vec{r}} = A$, see Fig. \ref{fig:SpatialMERA}(i) for an MPS. Translation invariance is in this way exactly preserved. It can also be exploited to reduce computational cost from $O(N)$ to $O(1)$. 

On the other hand, it is not clear how scale invariance could be enforced in these tensor networks. For an MPS this is unlikely to be possible at all because, as we mentioned earlier, a finite bond dimension $\chi$ already implies the presence of an effective finite correlation length $\zeta \approx \chi^\kappa$ \cite{Tagliacozzo08, Pollmann09}.

\begin{figure}[!bthp]
\begin{center}
\includegraphics[width=10cm]{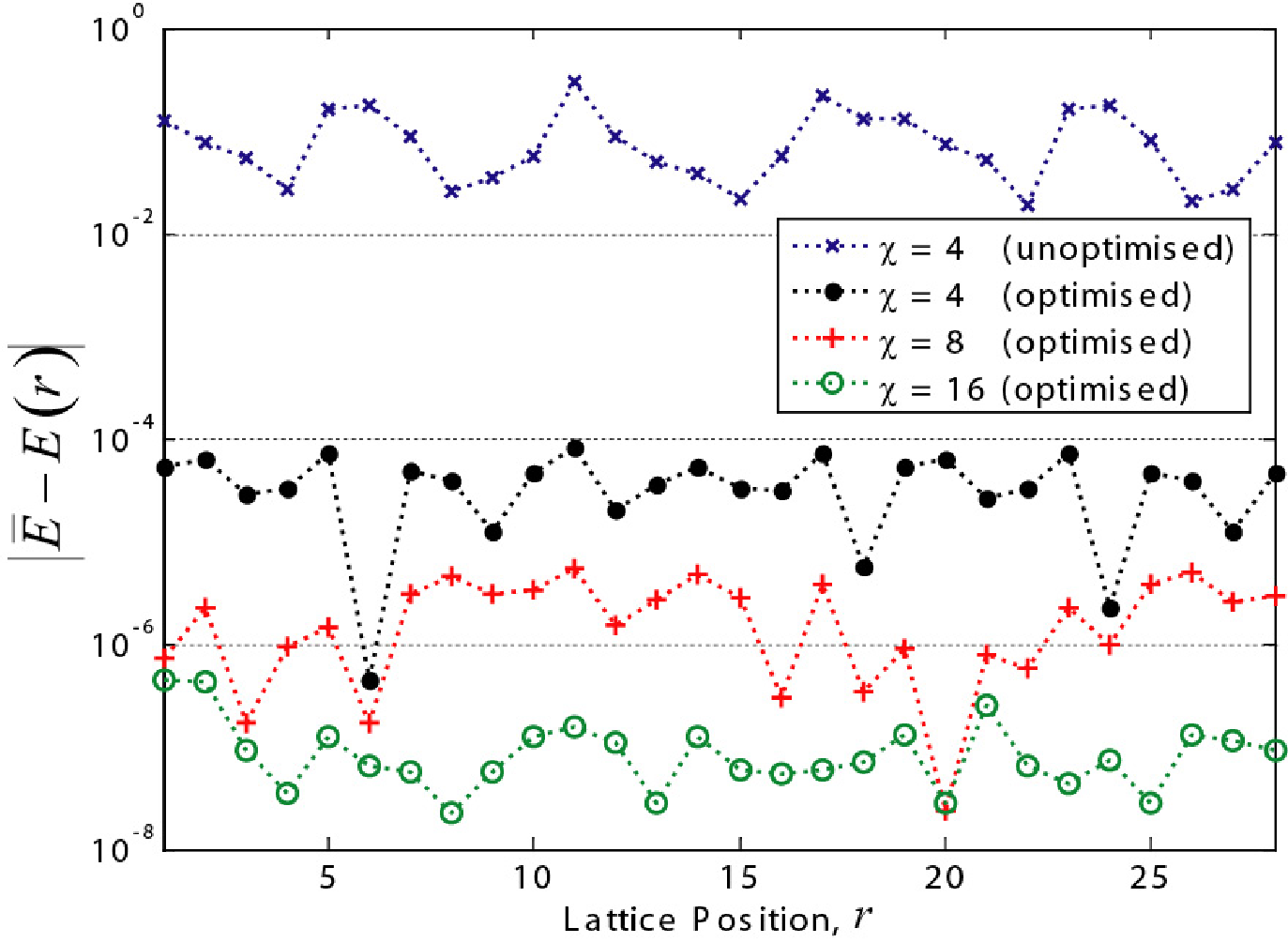}
\caption{We investigate translation invariance in the scale-invariant MERA by comparing the bond energy $E(r)$ over 30 contiguous lattice sites with the average bond energy $\bar E$ from all sites, as measured with the critical Ising Hamiltonian, $H_\textrm{Ising}$, of Eq. \ref{eq:sBe1}. For a randomly initialized $\chi=4$ scale-invariant MERA, the large fluctuations of bond energies indicate the state is highly non-translationally invariant. Once the MERA has been optimized for the ground state of $H_\textrm{Ising}$, it more closely approximates translation invariance; bond energies now differ from the average by less than $0.1 \%$. As the bond dimension $\chi$ of the MERA is increased, the optimized wavefunction better approximates translation invariance; for $\chi=16$ the bond energies differ from the average energy by less than $0.001 \%$.}
\label{fig:TransInvar}
\end{center}
\end{figure}

Let us now consider spatial symmetries in the MERA. Recall that a generic MERA on an $N$ site lattice is arranged into $T\approx \log(N)$ layers of tensors, and contains $O(N)$ different tensors, as depicted in Fig. \ref{fig:SpatialMERA}(iia). Suppose now that the state to be approximated by the MERA is translation invariant. Then we can choose all the tensors in each layer to be the same, so that layer $U^{[\tau]}$ is characterized by a single pair of tensors $u^{[\tau]}$ and $w^{[\tau]}$, see Fig. \ref{fig:SpatialMERA}(iib). In this way translation invariance can be exploited to reduce computational costs from $O(N)$ to $O(\log(N))$. Notice, however, that this choice of tensors does not enforce translation invariance, because the structure of the coarse-graining is not homogeneous (different sites are positioned in inequivalent positions with respect to the disentanglers and isometries). The final effect is examined in Fig. \ref{fig:TransInvar}. A MERA characterized by a single pair of tensors $u^{[\tau]}$ and $w^{[\tau]}$ for each layer, where these tensors are filled with random coefficients (compatible with the isometric constraints of Eq. \ref{eq:sRe3}), is highly non-translation invariant, with e.g. oscillations in the expectation value of the energy of the order of $0.1$ for the Hamiltonian $H_\textrm{Ising}$ of Eq. \ref{eq:sBe1}. Still, these violations of translation invariance decrease significantly once the tensors are optimized so as to minimize the expectation value of the translation invariant Hamiltonian $H_\textrm{Ising}$. Indeed, they become of order $10^{-5}$ for $\chi=4$ and decrease with increasing $\chi$. [In practice one can efficiently average the expectation value of a local observable over all possible lattice positions in order to further reduce the effect of these small violations of translation invariance]. We conclude that translation invariance can be exploited to reduce computational costs, but it can only be reproduced approximately. It is not known whether it can be enforced exactly.

Instead, enforcing scale invariance in the MERA is straightforward. This is accomplished by choosing all disentanglers and isometries to be copies of a single pair $u$ and $w$, see Fig. \ref{fig:SpatialMERA}(iic), which further reduces the number of parameters and the computational cost of MERA algorithms from $O(\log (N))$ to $O(1)$, allowing infinite systems to be considered. The scale-invariant MERA will be discussed in more detail Sect. \ref{sect:ScaleMERA}. 

To summarize, in the MPS/PEPS we can enforce and exploit translation invariance but not scale invariance, whereas in the MERA we can enforce and exploit scale invariance but only exploit (i.e., we cannot enforce) translation invariance. Thus both MPS/PEPS and MERA have potential advantages over each other, depending on whether exact translation invariance or exact scale invariance is more important for the problem under consideration.

\subsection{Global Internal Symmetries} \label{sect:SymInternal} 

A second important class of symmetries are those involving internal degrees of freedom, such as $\mathbb Z_2$ spin flips and $U(1)$ or $SU(2)$ spin rotations simultaneously applied on all the sites of a spin model. Such symmetries can be enforced and exploited in all tensor networks.

\begin{figure}[!htbp]
\begin{center}
\includegraphics[width=8cm]{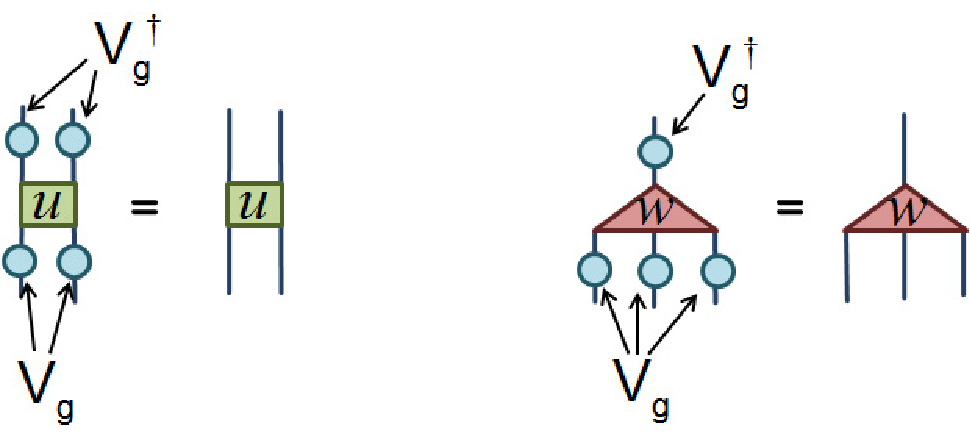}
\caption{In order to preserve a (global) symmetry specified by symmetry group $\mathcal{G}$, the tensors $u$ and $w$ comprising the MERA are chosen to be invariant under the action of a unitary representation $V_g$ of symmetry group $\mathcal{G}$, see also Eq. \ref{eq:sSe2}.}
\label{fig:NonLocCoarse}
\end{center}
\end{figure}

Let us assume that the Hamiltonian $H$ of our lattice model is invariant under a symmetry group $\mathcal{G}$,
\begin{equation}
 \Gamma_g ~H~ \Gamma_g^{\dagger} = H, ~~~~\forall g \in \mathcal{G}, \label{eq:sSe1}
\end{equation}
where $\Gamma_g \equiv \cdots V_g \otimes V_g \otimes V_g \cdots$ is an infinite string of copies of a matrix $V_g$, with $V_g$ a unitary representation of $\mathcal{G}$, and let $|\psi \rangle$ be the ground state of $H$, which we will assume to have the same symmetry, i.e. $\Gamma_g |{\psi}\rangle=|\psi\rangle$ (or, more generally, $\Gamma_g |{\psi}\rangle=e^{i\phi}|\psi\rangle$). We can then ensure that the symmetry is also exactly preserved in a tensor network approximation to $|{\psi}\rangle$ by using symmetry preserving tensors \cite{Singh10a,Singh10b}. For instance, for the MERA, we choose the disentanglers $u$ and isometries $w$ such that, 
\begin{eqnarray}
	(V_g \otimes V_g) ~u~ (V_g\otimes V_g)^{\dagger} &=& u, \nonumber\\
	(V_g \otimes V_g \otimes V_g) ~w~ (V_g)^{\dagger} &=& w \label{eq:sSe2},
\end{eqnarray}
where $V_g$ acting on different indices may actually denote different (in general, reducible) representations of $\mathcal{G}$, see also Fig. \ref{fig:NonLocCoarse}. The use of symmetry preserving tensors implies that the tensors are block diagonal when viewed in a certain basis and thus contain less free parameters than generic tensors. This reduction in the number of parameters can be exploited to significantly decrease computational costs.  Symmetries, and in particular a truncated version of the operator $\Gamma_g$, also play an important role in the description of non-local scaling operators, as discussed in Sect. \ref{sect:ScaleInvObjects}.

\section{Scale-invariant MERA} \label{sect:ScaleMERA} 

\begin{figure}[!htbp]
\begin{center}
\includegraphics[width=8cm]{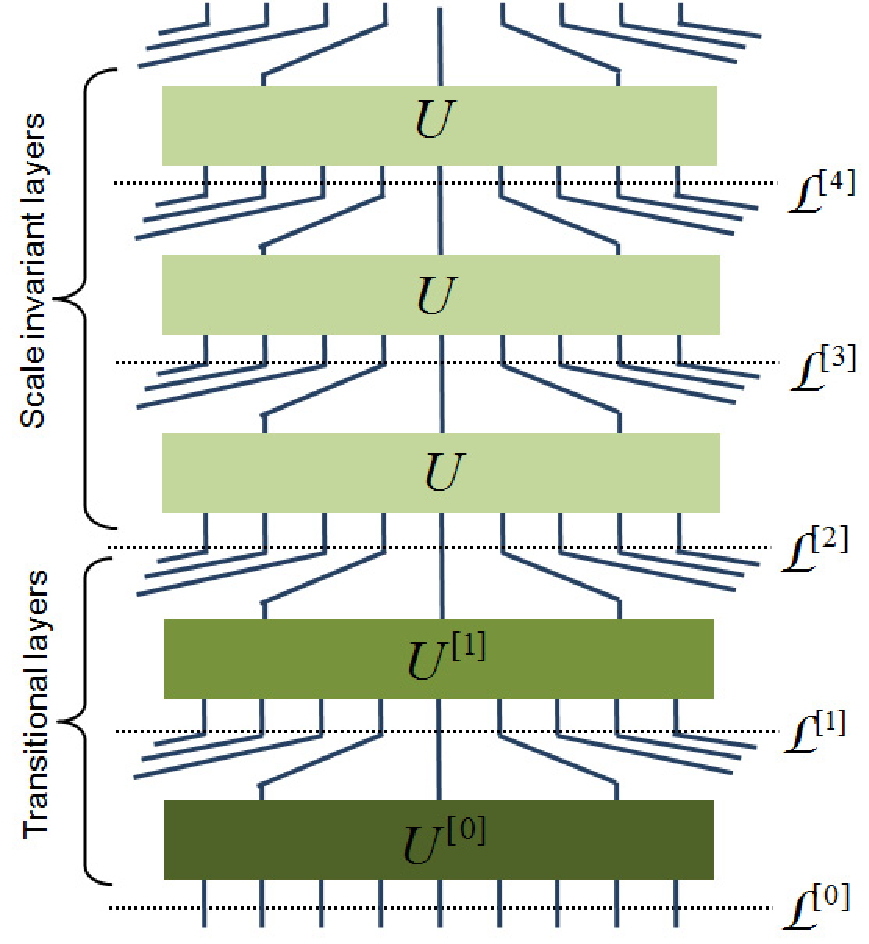}
\caption{A scale-invariant MERA consists of a number $M$ of transitional layers, here $M=2$ transitional layers $U^{[0]}$ and $U^{[1]}$, followed by an infinite number identical of scale-invariant layers $U$. Recall that each layer $U^{[\tau]}$ is comprised of local isometric tensors, the disentanglers $u^{[\tau]}$ and isometries $w^{[\tau]}$, as depicted Fig. \ref{fig:ERandMERA}(i).}
\label{fig:ScaleMERA}
\end{center}
\end{figure}

We have already introduced the scale-invariant MERA: in a lattice $\mathcal{L}$ with an infinite number of sites, $N\rightarrow \infty$, it consists of infinitely many layers of tensors, where all the disentanglers and isometries are copies of a unique pair $u$ and $w$. In this section we enumerate two significant structural properties of the scale-invariant MERA and review how one can compute a local reduced density matrix, from which the expectation value of a local operator can be evaluated. Then we discuss the three types of scale-invariant (or covariant) objects one can extract from it. 

\subsection{Basic Properties} \label{sect:BasicProp}

Two basic features of the scale-invariant MERA in $D=1$ dimensions match well-known properties of the ground state of a critical system. Firstly, the entanglement entropy $S_L$ of a block of $L$ contiguous sites can be seen to scale as the logarithm of $L$ \cite{Vidal08}, which is compatible with the critical scaling \cite{Vidal03b,Calabrese04},
\begin{equation}
	S_L\approx \frac{c}{3} \log (L),
\end{equation}
where $c$ is the central charge of the CFT. Secondly, correlation functions can be seen to decay polynomially \cite{Vidal08}, 
\begin{equation}
	\langle o(r_1) o(r_2) \rangle \approx \frac{1}{|r_1-r_2|^{q}},
\end{equation}
as it is also expected of critical correlators. Interestingly, these two properties of the scale-invariant MERA follow from simple geometric considerations, namely by studying minimally connected regions and geodesic paths in the (discrete) holographic geometry generated by the tensor network \cite{Evenbly11}.

\subsection{Transitional Layers}

In a practical computation (see Sect. \ref{sect:ScaleImp}) it is customary to consider a scale-invariant MERA with some small number $M$ of translational layers $\{U^{[0]}, \cdots, U^{[M-1]}\}$, which are characterized by $M$ pairs of tensors $\{(u^{[0]}, w^{[0]}),$ $\cdots,$ $(u^{[M-1]}, w^{[M-1]})\}$ that are chosen independently of the single pair $(u,w)$ characterizing the rest of layers (see Fig. \ref{fig:ScaleMERA}(i) for an example with $M=2$).  
These transitional layers serve two main purposes. Firstly, they allow one to choose the bond dimension $\chi$ of the scale-invariant layers independent of the local dimension $d$ of the sites in the original lattice $\mathcal{L}^{[0]}$.

Secondly they also allow to diminish the effect of RG irrelevant terms in the critical Hamiltonian $H$ of the system. Such terms violate scale invariance but become less and less important at larger length scales. The number M of transitional layers required depends on the amplitude and scaling dimensions of the irrelevant operators present in $H$, and is often determined by trial and error. For the sake of simplicity, in the rest of this section we shall focus on the case of a purely scale-invariant MERA with no transitional layers. 

\subsection{Local Density Matrix} \label{sect:LocDensity}

The computation of the (average) local density matrix 
\begin{equation}
\bar \rho^{}  \equiv \mathop {\lim }\limits_{N \rightarrow \infty } \left( {\frac{1}{N}\sum\limits_{r = 1}^N {\rho^{} (r ,r+1 )} } \right)
\label{eq:rhoAv}
\end{equation}
for two contiguous sites of lattice $\mathcal{L}$ is of central importance in the present formalism. The density matrix $\bar{\rho}$ is required both in order to extract the expectation value of a local operator $o$ from the scale-invariant MERA and to optimize its tensors so as to approximate the ground state of a critical Hamiltonian $H$.

Here we consider the evaluation of the expectation value $\left\langle {o(r, r+1)} \right\rangle$ of a local observable $o(r, r+1 )$. As discussed in Sect. \ref{sect:SymSpatial}, the MERA in not manifestly translation invariant. Thus the expectation value $\left\langle {o(r, r+1)} \right\rangle$ can \textit{artificially} vary with the position of site $r$ in the lattice. To mitigate this effect, rather than evaluating the expectation value at a particular lattice position $r$ we will instead evaluate the average expectation value over all lattice sites,
\begin{equation}
\left\langle {\bar o} \right\rangle  \equiv \mathop {\lim }\limits_{N \rightarrow \infty } \left( \frac{1}{N}\sum\limits_{r = 1}^N {\left\langle {o(r, r+1)} \right\rangle }\right). \label{eq:sRe13b}
\end{equation}
Notice that this average expectation value can be expressed in terms of the average two-site reduced density matrix $\bar \rho$ introduced in Eq. \ref{eq:rhoAv},
\begin{equation}
\left\langle {\bar o} \right\rangle  = \textrm{tr} \left(~{o(r,r+1) ~\bar{ \rho}~} \right).
\end{equation}

In Sect. \ref{sect:MERAfoundation} we described the use of the left, center and right descending superoperators, $\left\{ \mathcal{D}_L, \mathcal{D}_C, \mathcal{D}_R \right\}$, to compute the reduced density matrix $\rho^{[0]}(r,r+1)$ from a finite MERA. In particular, it was argued that obtaining the density matrix $\rho^{[0]}(r,r+1)$ required application of a specific sequence of left, center and right descending superoperators that depended on the causal cone associated to sites $(r,r+1)$, see Eq. \ref{eq:sRe12} for an example. The average density matrix $\bar \rho$ can be seen to follow from using the average descending superoperator $\mathcal{\bar D}$, defined as
\begin{equation}
\mathcal{\bar D} \equiv \frac{1}{3} \left( \mathcal D_L + \mathcal D_C + \mathcal D_R \right), \label{eq:sRe14}
\end{equation}
in order to descend the density matrix through the `average' causal cone. That is, given the average density matrix $\bar \rho^{[\tau]}$ at level $\tau$, the average density matrix $\bar \rho^{[\tau-1]}$ at lower level $\tau-1$ is obtained as
\begin{equation}
\bar \rho^{[\tau-1]} = {{\cal {\bar D}}} \left( \bar \rho^{[\tau]} \right). \label{eq:sRe15}
\end{equation}

In an infinite system, $N\rightarrow \infty$, the MERA has $T\rightarrow \infty$ layers, and the average density matrix $\bar \rho$ is obtained from
\begin{equation}
 \bar \rho = \lim_{T\rightarrow \infty} \big(\underbrace{\mathcal{\bar D}\circ \cdots \circ \mathcal{\bar D}}_{T \mbox{ \scriptsize{times}}}\big) (\bar \rho^{[T]}), \label{eq:sRe16}
\end{equation}
where $\bar \rho$ is simply the dominant eigenoperator of the descending superoperator $\mathcal{\bar D}$, which is independent of ${\bar \rho^{[T]}}$. As a manifestation of scale invariance, this is also the two-site density matrix of any coarse-grained lattice $\mathcal{L}^{[\tau]}$, that is ${\bar \rho^{[\tau]}}={\bar \rho}$ for any $\tau\geq 0$. More details on the computation of $\bar \rho$ can be found in Sect. \ref{sect:ScaleImp}.

\begin{figure}[!htbp]
\begin{center}
\includegraphics[width=12cm]{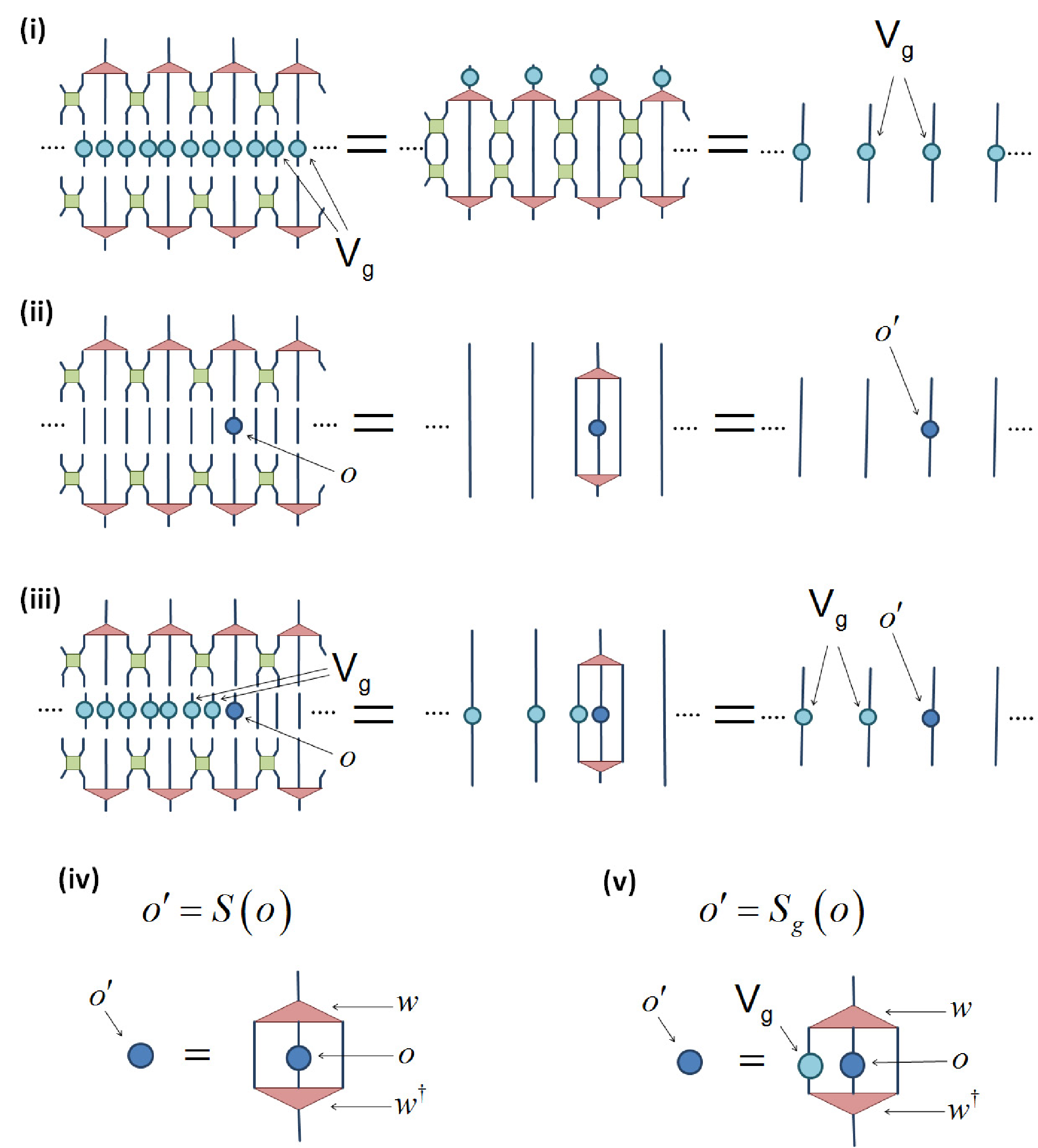}
\caption{The three classes of scale-invariant/covariant objects for a $D=1$ dimensional lattice include (i) an infinite string of $V_g$, with $V_g$ a unitary representation of symmetry group $\mathcal{G}$ of the system, (ii) local operators and (iii) non-local operators, which consist of a local operator with a semi-infinite `tail' of $V_g$. (iv) Scaling superoperator $\mathcal S$ for local operators and (v) scaling superoperator $\mathcal S_g$ for non-local operators.}
\label{fig:ScaleObject}
\end{center}
\end{figure}

\subsection{Scale-invariant Objects} \label{sect:ScaleInvObjects}

The scale-invariant MERA offers direct access to objects of a critical theory that are invariant (more generally, covariant) under a change of scale. In $D=1$ dimensions we can identify three classes of such objects, as depicted Fig. \ref{fig:ScaleObject}(i-iii). Next we discuss them in some detail.

\subsubsection{Symmetry Transformations} 
Let us assume that the critical ground state is invariant under some symmetry group $\mathcal{G}$, as implemented by the symmetry transformations $\Gamma_g \equiv \cdots V_g \otimes V_g \otimes V_g \cdots$ introduced in Eq. \ref{eq:sSe1}, where $g \in \mathcal{G}$. The first type of objects that transform into themselves under changes of scale correspond precisely to the infinite strings $\Gamma_g$, for which we have
\begin{equation}
	\Gamma_g ~~ \stackrel{U^{\dagger}}{\longrightarrow} ~~ \Gamma_g
\end{equation}
 
Indeed, if we use symmetry preserving tensors in the MERA, as per Eq. \ref{eq:sSe2}, then the string $\Gamma_g$ commutes with each layer $U$ of the MERA, $\Gamma _g U = U\Gamma _g$, 
or equivalently the string $\Gamma_g$ remains invariant under coarse-graining, $U^\dag \Gamma _g U = \Gamma _g$, as shown in Fig. \ref{fig:ScaleObject}(i).

\subsubsection{Local Scaling Operators} 

The second class of objects with a simple transformation rule under coarse-graining, which can be easily extracted from the scale-invariant MERA, are local scaling operators $\phi_{\alpha}$, fulfilling
\begin{equation}
		\phi_{\alpha} ~~ \stackrel{U^{\dagger}}{\longrightarrow} ~~ \lambda_{\alpha}	~ \phi_{\alpha},
\end{equation}
where $\lambda_{\alpha}$ is some constant.

For simplicity, here we focus on one-site scaling operators [One can also compute two-site scaling operators, but they lead to the same scaling dimensions and fusion rules]. As depicted in Fig. \ref{fig:ScaleObject}(ii), a one-site operator $o$ located at certain points on the lattice is coarse-grained into another one-site operator $o'$.  This coarse-graining can directly be implemented with the one-site ascending superoperator, which we call the (one-site) scaling superoperator $\mathcal S$ in the scale-invariant setting,
\begin{equation}
o'=\mathcal S (o),
\end{equation}
see also Fig. \ref{fig:ScaleObject}(iv). Iteration produces an RG flow for one-site operators:
\begin{equation}
o  \stackrel{\mathcal{S} }{\longrightarrow}  o' \stackrel{\mathcal{S}}{\longrightarrow} o''  ~\cdots~.
\label{eq:sCe3b}
\end{equation}
The scaling operators $\phi_\alpha$ and their corresponding scaling dimensions $\Delta_\alpha$,
\begin{equation}
	\mathcal{S}(\phi_{\alpha}) = \lambda_{\alpha} \phi_{\alpha},~~~~~~~\Delta_{\alpha} \equiv -\log_3 \lambda_{\alpha}, \label{eq:sCe1}
\end{equation}
can then be obtained by simply diagonalizing the scaling superoperator $\mathcal{S}$ \cite{Giovannetti08,Pfeifer09}.

\subsubsection{Non-Local Scaling Operators} 

Let us assume again that the critical ground state represented by the scale-invariant MERA is invariant under a symmetry group $\mathcal{G}$, as implemented by the symmetry transformations $\Gamma_g \equiv \cdots V_g \otimes V_g \otimes V_g \cdots$ introduced in Eq. \ref{eq:sSe1}, where $g \in \mathcal{G}$, and that the tensors of the MERA have been chosen to preserve this symmetry, as per Eq. \ref{eq:sSe2}. We can then identify a third class of objects with a simple transformation rule under changes of scale, namely non-local scaling operators $\phi^{\triangleleft}_{g,\alpha}$, to be defined below, which fulfill
\begin{equation}
		\phi^{\triangleleft}_{g,\alpha} ~~ \stackrel{U^{\dagger}}{\longrightarrow} ~~ \lambda_{g,\alpha}~	\phi^{\triangleleft}_{g,\alpha},
\end{equation}
where $\lambda_{g,\alpha}$ is some constant.

To see how these scaling operators come about \cite{Evenbly10d}, let us first introduce non-local operators $o^{\triangleleft}_g$ of the form,
\begin{equation}
	o^{\triangleleft}_g = \Gamma^{\triangleleft}_g \otimes o, ~~~~~ \Gamma^{\triangleleft}_g \equiv  \underbrace{\cdots V_g \otimes V_g \otimes V_g}_{\infty} \label{eq:sCe7}
\end{equation}
where $\Gamma^{\triangleleft}_g$ is a semi-infinite string made of copies of $V_g$ and $o$ is a one-site operator attached to the open end of $\Gamma^{\triangleleft}_g$. Notice that, under coarse-graining, $o^{\triangleleft}_g$ can be mapped into another non-local operator ${o^{\triangleleft}_g}'$ of the same type,
\begin{equation}
	o^{\triangleleft}_g = \Gamma^{\triangleleft}_g \otimes o ~~~\stackrel{U^{\dagger}}{\longrightarrow} ~~~ {o^{\triangleleft}_g}' = \Gamma^{\triangleleft}_g \otimes o', \label{eq:sCe8}
\end{equation}
since the semi-infinite string $\Gamma_{g}$ commutes with the coarse-graining everywhere except at its open end, as illustrated in Fig. \ref{fig:ScaleObject}(iii). Thus we can study the sequence of coarse-grained non-local operators $o^{\triangleleft}_g \longrightarrow {o^{\triangleleft}_g}' \longrightarrow {o^{\triangleleft}_g}'' \cdots$ by just coarse-graining the local operator $o$ with the modified one-site scaling superoperator $\mathcal{S}_{g}$ of Fig. \ref{fig:ScaleObject}(v),
\begin{equation}
o  \stackrel{\mathcal{S}_{g}}{\longrightarrow}  o' \stackrel{\mathcal{S}_{g}}{\longrightarrow} o''  ~\cdots~.
\label{eq:sCe9}
\end{equation}
In particular we can diagonalize the modified scaling superoperator $\mathcal{S}_{g}$,
\begin{equation}
	\mathcal{S}_{g}(\phi_{g,\alpha}) = \lambda_{g,\alpha} \phi_{g,\alpha}~, ~~~~~\Delta_{g,\alpha} \equiv -\log_3 \lambda_{g,\alpha}~,	\label{eq:sCe10}
\end{equation}
to obtain non-local scaling operators $\phi^{\triangleleft}_{g,\alpha}$ of the form
\begin{equation}
	\phi^{\triangleleft}_{g,\alpha} = \Gamma^{\triangleleft}_g \otimes \phi_{g,\alpha}. \label{eq:sCe11}
\end{equation}
Notice that for $g = \mathbb{I}$ we recover the local scaling operators $\phi_{\alpha}$ of Eq. \ref{eq:sCe1}.

Importantly, the scaling dimensions $\Delta_{\alpha}$ and $\Delta_{g,\alpha}$ of the both local and non-local scaling operators $\phi_{\alpha}$ and $\phi^{\triangleleft}_{g,\alpha}$ (as well as their operator product expansion coefficients, see Ref.\onlinecite{Pfeifer09}) are the same in the lattice than in the continuum. Therefore by extracting properties of the scaling operators on the lattice, we can characterize the CFT that describes the critical theory in the continuum. As demonstrated by the benchmark results of Sect. \ref{sect:CFTBench}, a relatively simple and inexpensive MERA simulation can actually be used to obtain remarkably accurate conformal data of the underlying CFT.

\section{Benchmark Results} \label{sect:MERABench}

In this section we benchmark the performance of the scale-invariant MERA by applying it to study of the ground state of several well-known quantum spin chains at criticality. The models we analyze are the critical Ising model \cite{Pfeuty70,Burkhardt85}, the critical three-state Potts models \cite{Solyom81}, the XX model \cite{Lieb61} and a Heisenberg zig-zag chain \cite{Eggert96} (the Heisenberg model with a next-nearest neighbor coupling), corresponding to the following Hamiltonians: 
\begin{align}
 H_{{\rm{Ising}}}  &= \sum\limits_r {\Big( {Z{(r)}  - X{(r)} X{(r + 1)} } \Big)} \label{eq:sBe1} \\ 
 H_{{\rm{Potts}}}  &= \sum\limits_r {\left( {\tilde Z{(r)}  - \tilde X {(r)} \tilde X^{\dagger} {(r + 1)} -\tilde X^{\dagger} {(r)}\tilde X {(r + 1)} } \right)} \label{eq:sBe2} \\ 
 H_{{\rm{XX}}}     &= \sum\limits_r {\Big( {X{(r)} X{(r + 1)}  + Y{(r)} Y{(r + 1)} } \Big)} \label{eq:sBe3} \\ 
 H_{{\rm{Heis.Zig-Zag}}} & = \sum\limits_r {\left( {\vec{S}(r) \cdot \vec S(r + 1) + J_2 \vec S(r) \cdot \vec S(r + 2)} \right)}   \label{eq:sBe4}
\end{align}
where $X$, $Y$, $Z$ are Pauli matrices, $\vec S = \left[ {X,Y,Z} \right]$, and where $\tilde Z$, $\tilde X$ are three-state Potts spin matrices given by
\begin{equation}
\tilde Z  \equiv \left( {\begin{array}{*{20}c}
   -1 & 0 & 0  \\
   0 & { 2} & 0  \\
   0 & 0 & { - 1}  \\
\end{array}} \right),\; \; 
\tilde X  \equiv \left( {\begin{array}{*{20}c}
   0 & 1 & 0  \\
   0 & 0 & 1  \\
   1 & 0 & 0  \\
\end{array}} \right).\label{eq:sBe5} 
\end{equation}
The next-nearest neighbor coupling in the Heisenberg zig-zag chain is set at the critical value $J_2=0.24116$; at this value the model is scale-invariant \cite{Eggert96}. Note that, although the standard Heisenberg model (with $J_2=0$) is quantum critical, the Hamiltonian contains a marginally irrelevant contribution that breaks scale invariance, which here we remove by adding the next-nearest neighbor coupling. 

In the present calculation, we have used the modified binary MERA scheme, depicted in Fig. \ref{fig:SchemeMERA}(iii), with either $M=2$ or $3$ transition layers. This ansatz is optimized by minimizing the expectation value of the energy density for each of the above Hamiltonians, by using the optimization algorithm described in Sect. \ref{sect:ScaleImp}. In terms of the bond dimension $\chi_{\mbox{\tiny MERA}}$, the cost of optimizing the modified binary MERA scales as $O(\chi_{\mbox{\tiny MERA}}^7)$, but can be reduced to $O(\chi_{\mbox{\tiny MERA}}^6)$ through use of an approximation in the tensor network contractions as described Sect. \ref{MERAapprox}. We have computed the ground states of the four models over a range of values of $\chi_{\mbox{\tiny MERA}}$ up to $\chi_{\mbox{\tiny MERA}} = 150$. Each simulation took under a week on a 3GHz dual-core workstation with 32Gb of RAM. 

For purposes of comparison, we have also computed the ground state of the four critical spin chains using an infinite, translation invariant (with a 4-site unit cell) MPS. The MPS tensors are optimized with a variational approach similar to the iDMRG algorithm \cite{McCulloch08}. The computational cost scales with the bond dimension $\chi_{\mbox{\tiny MPS}}$ of the MPS as $O(\chi_{\mbox{\tiny MPS}}^3)$. We have computed the ground states of the four models over a range of values of $\chi_{\mbox{\tiny MPS}}$ up to $\chi_{\mbox{\tiny MPS}} = 1536$. 

In both the MERA and MPS calculations we have employed symmetry preserving tensors, as described in Sect. \ref{sect:SymInternal}, to enforce (some of) the global internal symmetries of these models. Specifically, $\mathbb Z_2$ symmetric tensors have been used for the Ising model; $\mathbb Z_3$ symmetric tensors have been used for the Potts model ($\mathbb Z_3$ is a subgroup of the full $S_3$ symmetry of this model); and $U(1)$ symmetric tensors have been used for both the quantum XX and Heisenberg zig-zag chains (again, $U(1)$ is a subgroup of the full $SU(2)$ symmetry of the Heisenberg zig-zag chain). 

In the first part of the benchmark, Sect. \ref{sect:MERAvsMPS}, we compare ground energy and two-point correlators obtained from MERA and MPS, and discuss the relative merits of each approach. Then in Sect. \ref{sect:CFTBench} we demonstrate the extraction of conformal data from the scale-invariant MERA for the critical Ising model, following the approaches of Refs.\cite{Pfeifer09,Evenbly10d}.

\subsection{Comparison with MPS} \label{sect:MERAvsMPS}

Here we compare the performances of MPS and scale-invariant MERA for the computation of ground state energy and two-point correlators.

\subsubsection{Ground Energy} \label{sect:EnergyCompare}

\begin{figure}[!tbp]
\begin{center}
\includegraphics[width=10cm]{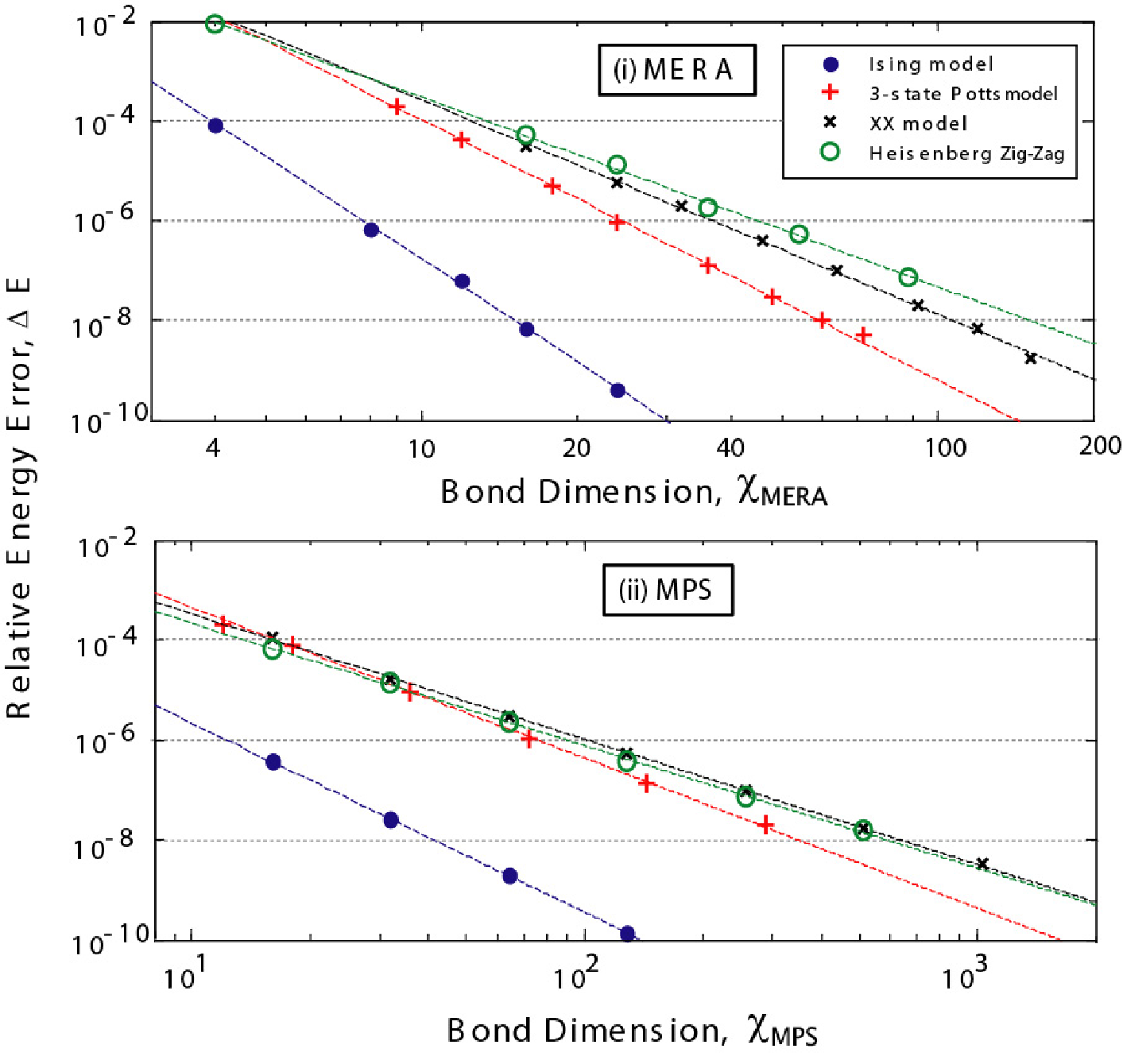}
\caption{Relative energy error $\Delta E$ in the ground state of critical spin models as a function of the tensor network bond dimension $\chi$, comparing (i) the scale-invariant MERA with (ii) an infinite MPS. In all cases the energy error $\Delta E$ appears to scale polynomially in $\chi$ in accordance with Eq. \ref{eq:sBe6}. }
\label{fig:MERAandMPSenergy}
\end{center}
\end{figure}

For both Ising and quantum XX models, the exact ground energy per site is $E=-4/\pi$, while for the three-state Potts and Heisenberg zig-zag chains we use an MPS with $\chi_{\mbox{\tiny MPS}}=1536$ to estimate the ground energy per site at $E_\textrm{Potts}= -2.4359911239(1)$ and $E_\textrm{Heis.Z.Z.} \approx -1.607784749(1)$. 

Fig. \ref{fig:MERAandMPSenergy} displays the relative error in the ground state energy per site, $\Delta E \equiv (E_{\textrm{exact}} - E_{\textrm{numeric}})/ E_{\textrm{exact}}$ for the models under consideration over a range of bond dimensions $\chi$, for both MERA and MPS. This figure reveals a number of interesting similarities between results obtained with MERA and MPS. Recall that the central charge $c$ for these models is
\begin{equation}
	c_\textrm{Ising} = \frac{1}{2},~~~~c_\textrm{Potts} = \frac{4}{5}, 
	~~~~ c_\textrm{XX} = 1, ~~~~ c_\textrm{Heis.Z.Z.} = 1,
\end{equation}
Then a first observation is that for both MERA and MPS, for a given bond dimension $\chi$ the larger the central charge $c$ the larger the error in the energy is. A second similarity is that for both MERA and MPS the error $\Delta E$ in the energy scales polynomially in $\chi$, i.e. to a good approximation,
\begin{equation}
\Delta E = a\chi ^{ - b}. \label{eq:sBe6}
\end{equation}
A linear fit in Fig. \ref{fig:MERAandMPSenergy} produced the estimates for the coefficients $a$ and $b$ displayed in Table \ref{tab:FitCoeff}. In the large $\chi$ regime, the error $\Delta E$ is dominated by the coefficients $b$. Interestingly, the ratio $b_\textrm{MERA}/ b_\textrm{MPS}$ for the four models produces very similar results, namely $1.80, 1.74, 1.72,$ and  $1.56$ for the Ising, Potts, XX and Heisenberg zig-zag models respectively. Given that the average ratio is $b_\textrm{MERA}/ b_\textrm{MPS}\approx 1.72$, we conclude that in the large $\chi$ limit a similar error in the energy for MERA and MPS is obtained if
\begin{equation}
\Delta E_{{\rm{MERA}}} \left( \chi  \right) \approx \Delta E_{{\rm{MPS}}} \left( {\chi ^{1.72} } \right), \label{eq:sBe7b}
\end{equation}
that is, if $\chi_{\mbox{\tiny MPS}} = (\chi_{\mbox{\tiny MERA}})^{1.72}$. Taking into account that the number of variational parameters in the MERA and MPS scale as $(\chi_{\mbox{\tiny MERA}})^4$ and $(\chi_{\mbox{\tiny MPS}})^2$, this comparison shows that in the large $\chi$ limit the MPS requires less variational parameters than the scale-invariant MERA in order to obtain a similar accuracy in the ground state energy.

\begin{table}[!btp]
\centering
\caption[]{Best fit coefficients to the functional form of Eq. \ref{eq:sBe6} for the scaling of relative energy error in ground state MERA and MPS calculations. The central charge $c$ of the critical models is given for reference.}
\renewcommand{\arraystretch}{1.2}
\setlength\tabcolsep{5pt}
\begin{tabular}{ |c | r r | r r |}
\hline
$~~~~~$  & \multicolumn{2}{c|}{(i) MERA} & \multicolumn{2}{c|}{(ii) MPS} \\
$~~~~~$  & $a~~$ & $b~~$ & $a~~$ & $b~~$  \\
\hline
Ising Model ($c=1/2$)             & $~~$ 1.13       & 6.81  &  $~$ 0.013       & 3.78       \\
Potts Model ($c=4/5$)             & $~~$ 17.80      & 5.22  &  $~$ 0.432       & 3.00       \\
XX Model ($c=1$)                  & $~~$ 5.25       & 4.30  &  $~$ 0.103       & 2.50       \\
Heisenberg Zig-Zag ($c=1$)        & $~~$ 1.89       & 3.80  &  $~$ 0.059       & 2.44       \\
\hline
\end{tabular}
\label{tab:FitCoeff}       
\end{table}

It is tempting to extend this comparison to computational costs. A first step in this direction is to note that each iteration in the optimization of MERA and MPS scales (naively) as $(\chi_{\mbox{\tiny MERA}})^6$ and $(\chi_{\mbox{\tiny MPS}})^3$, from which it would be tempting to conclude that MPS algorithms require a lower computational budget than MERA algorithms to obtain the same accuracy in the ground state energy. However, there are important multiplicative prefactors $k_\textrm{MERA}$ and $k_\textrm{MPS}$ modifying the naive scaling of costs. In both cases, one is required to find the dominant eigenvector of a transfer matrix. But while in the case of the MERA this transfer matrix (or scaling superoperator) has a well-defined gap, implying that $k_\textrm{MERA}$ is essentially independent of $\chi_{\mbox{\tiny MERA}}$, in the case of the MPS the gap in the transfer matrix closes to zero with increasing $\chi_{\mbox{\tiny MPS}}$, and the prefactor $k_\textrm{MPS}$ actually also grows with growing $\chi_{\mbox{\tiny MPS}}$. Therefore a proper comparison of computational costs requires first a careful characterization of the dependence of $k_\textrm{MPS}$ in $\chi_{\mbox{\tiny MPS}}$, which is beyond the scope of the present manuscript.
 
\subsubsection{Two-Point Correlators} \label{sect:CorrCompare}

Let us now compare the accuracy of two-point correlators produced by the scale-invariant MERA and MPS. For both approaches, we consider the ground state of the quantum XX model and compute the correlator,
\begin{equation}
C\left( d \right) \equiv \left\langle {\hat a_{}^\dag  (r)\hat a(r + d)} \right\rangle \label{eq:sOe24}
\end{equation}
where $\hat a$ is a fermionic operator defined in terms of spin operators as
\begin{equation}
\hat a\left( r \right) = \left( {\prod\limits_{m < r} { Z(m)} } \right)\frac{{ X(r) - i Y(r)}}{2}. \label{eq:sOe25}
\end{equation}
The correlation function of Eq. \ref{eq:sOe24} has the exact expression
\begin{equation}
C_{{\rm{exact}}} \left( d \right) = \frac{{ - \sin \left( {\pi d /2} \right)}}{{\pi d}}, \label{eq:sOe26}
\end{equation}
which is obtained from mapping the quantum XX model to a free-fermion model \cite{Rico04}. The correlation function $C(d)$ decays polynomially, as expected from the ground state of a quantum critical model.

\begin{figure}[!htbp]
\begin{center}
\includegraphics[width=10cm]{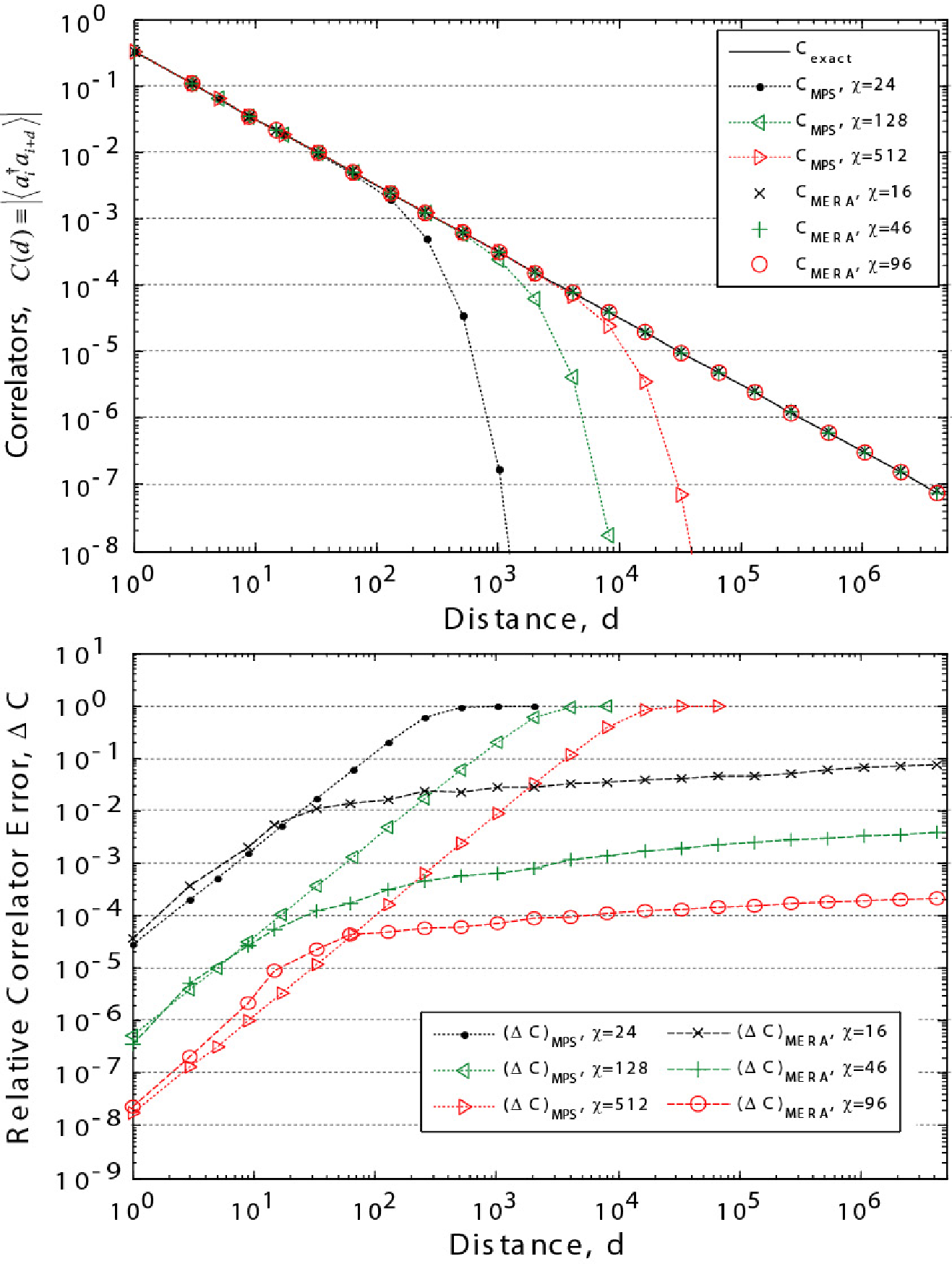}
\caption{(i) The two-point correlator of fermion operators, as defined Eq. \ref{eq:sOe24}, in the ground state of the quantum XX model, comparing results from the scale-invariant MERA and infinite MPS with the exact correlators. Correlators from MPS approximate polynomial decay only until some finite length scale $\zeta\approx\chi^{1.38}$, while correlators from MERA remain polynomial at all length scales. (ii) Relative error in correlators as defined Eq. \ref{eq:sOe28}.}
\label{fig:CorrErr}
\end{center}
\end{figure}

Fig. \ref{fig:CorrErr}(i) shows the correlations obtained with a scale-invariant MERA for $\chi_{\mbox{\tiny MERA}}=\{ 16, 46, 96\}$ and with an MPS for $\chi_{\mbox{\tiny MPS}}=\{ 24, 128, 512 \}$. These particular values of the bond dimension $\chi$ have been chosen so that the tensor networks produce comparable errors in the ground state energy, namely
\begin{align}
\Delta E_{{\rm{MERA}}} \left( \chi_{\mbox{\tiny MERA}} = 16 \right) & \approx \Delta E_{{\rm{MPS}}} \left( \chi_{\mbox{\tiny MPS}} = 24 \right) \approx 3.5\times 10^{-5} \nonumber \\
\Delta E_{{\rm{MERA}}} \left( \chi_{\mbox{\tiny MERA}} = 46 \right) & \approx \Delta E_{{\rm{MPS}}} \left( \chi_{\mbox{\tiny MPS}} = 128 \right) \approx 4.5\times 10^{-7} \nonumber \\
\Delta E_{{\rm{MERA}}} \left( \chi_{\mbox{\tiny MERA}} = 96 \right) & \approx \Delta E_{{\rm{MPS}}} \left( \chi_{\mbox{\tiny MPS}} = 512 \right) \approx 1.7\times 10^{-8}. \label{eq:sOe27}
\end{align}
The figure illustrates quite clearly that while the MPS can only approximate polynomial correlations up to a finite length scale $\zeta=\chi^\kappa$ \cite{Tagliacozzo08}, with $\kappa\approx 1.38$ for the quantum XX model, correlations in the MERA decay polynomially at all length scales \cite{Vidal08,Pfeifer09}. Fig. \ref{fig:CorrErr}(ii) displays the relative error in correlators, 
\begin{equation}
\Delta C = \frac{{\left| {C_{{\rm{exact}}}  - C_{{\rm{numeric}}} } \right|}}{{\left| {C_{{\rm{exact}}} } \right|}}. \label{eq:sOe28}
\end{equation}
Here it is seen that a MERA and an MPS that produce the same accuracy in the ground state energy produce also similarly accurate correlators at short distances, but the relative error in the correlator grows much slower with distance in the case of the MERA. For instance, the $\chi=96$ MERA reproduces correlators up to $d=10^6$ sites with relative error $\Delta C < 2\times 10^{-4}$, whereas the $\chi=512$ MPS, although possessing a similar ground energy, reproduces similarly accurate correlators only up to $d\approx100$ sites. 

\subsubsection{Summary of Comparison} \label{sect:ConcluCompare}

To summarize, we have seen that the MPS is more efficient than the scale-invariant MERA when it comes to ground state energies of critical Hamiltonians, in that it requires less variational parameters to achieve the same accuracy. However, the MERA produces better correlators at large distances, and it is therefore better suited to characterize asymptotic behaviors, such as the polynomial decay of correlations, from which one could in principle extract the critical exponents of the theory. However, as exemplified in Sect. \ref{sect:ScaleInvObjects} with the computation of scaling dimensions for local and non-local operators, critical exponents and other conformal data can actually be extracted more directly by analyzing the scaling superoperator. We illustrate this next.

\subsection{Evaluation of Conformal Data: The Ising Model} \label{sect:CFTBench}

As an example of extraction of conformal data from the scale-invariant MERA, here we identify the whole operator content (local and non-local primary fields) of the CFT corresponding to the quantum critical Ising model, reproducing the analysis of Ref.\onlinecite{Evenbly10d}. Similar results have also been obtained for the three-state Potts and quantum XX models in Refs. \cite{Pfeifer09, Evenbly10d}. 

The Hamiltonian $H_\textrm{Ising}$ of Eq. \ref{eq:sBe1} has a global internal $\mathbb{Z}_2$ corresponding to flipping all the spins. That is, $\mathcal{G} = \mathbb{Z}_2$ and $g\in \{+1,-1\}$, with $V_{+1} = \mathbb{I}$ and $V_{-1} = Z$, and
\begin{equation}
	\Gamma_{-1} ~H_\textrm{Ising}~ \Gamma_{-1}^{\dagger} = H_\textrm{Ising},~~~~~\Gamma_{-1} \equiv \bigotimes_{m=-\infty}^{\infty} Z. \label{eq:sCe22}
\end{equation}
The tensors $u$ and $w$ that comprise the scale-invariant MERA are chosen to be parity preserving; each index $i$ of tensors $u$ and $w$ decomposes as $i=(p,\alpha_p)$, where $p$ labels the parity ($p=+1$ for even parity and $p=-1$ for odd parity) and $\alpha_p$ labels the distinct values of $i$ with parity $p$. For tensors $u$, $w$ to be parity preserving it is ensured that, e.g. $u_{i_1,i_2}^{j_1,j_2} = 0$ if $p(i_1)p(i_2)p(j_1)p(j_2)=-1$, in accordance with Eq. \ref{eq:sSe2}. An operator $O$ acting on the spin chain has parity $p$ if
\begin{equation}
(\Gamma_{-1}) ~ O ~ (\Gamma_{-1})^{\dagger} = p ~ O. \label{eq:sCe23}
\end{equation} 
The local scaling superoperator $\mathcal{S}_{g=1}$ and the non-local scaling superoperator $\mathcal{S}_{g=-1}$, see Fig. \ref{fig:ScaleObject}(iv-v), are obtained from the optimized scale-invariant MERA with bond dimension $\chi=32$. The scaling superoperator $\mathcal{S}_{g=1}$ is diagonalized to find the local scaling operators $\phi_{+1,\alpha}$ together with their scaling dimensions $\Delta_{+1,\alpha}$, while the scaling superoperator $\mathcal{S}_{g=-1}$ is diagonalized to find the non-local scaling operators of the form
\begin{equation}
	\phi^{\triangleleft}_{-1,\alpha} = \cdots Z\otimes Z \otimes Z \otimes \phi_{-1,\alpha}, \label{eq:sCe24}
\end{equation}
together with their scaling dimensions $\Delta_{-1,\alpha}$. 

\begin{figure}[!htbp]
\begin{center}
\includegraphics[width=8cm]{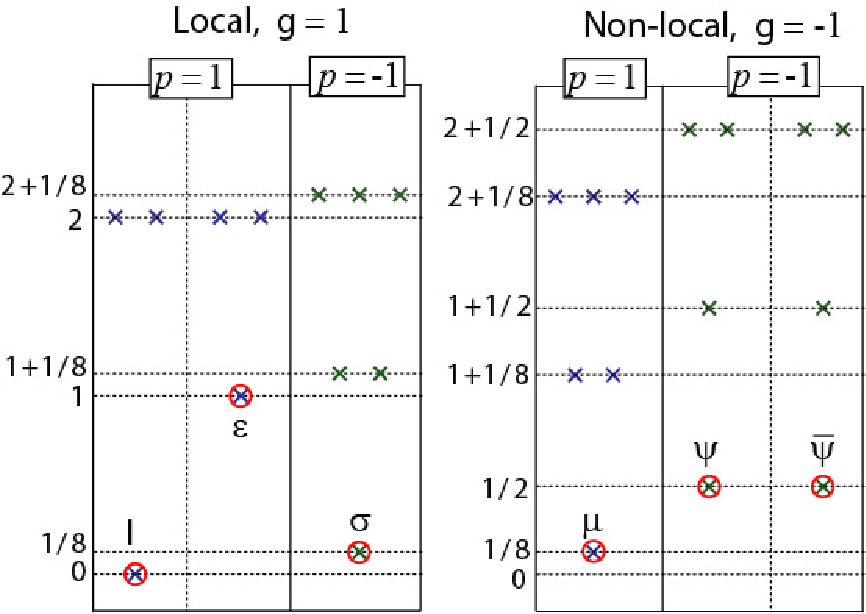}
\caption{A few scaling dimensions of the critical Ising model obtained from a $\chi=32$ scale-invariant MERA. The scaling dimensions are organized by the both locality $g=\pm1$ (local/non-local) and parity $p=\pm1$ (even/odd) of the corresponding scaling operators and together form the six conformal towers of the Ising CFT.}
\label{fig:IsingCritExp}
\end{center}
\end{figure}

Table \ref{tab:IsingExp} compares the exact scaling dimensions of the primary fields of the Ising CFT and their numerical estimates obtained from the MERA, which reproduce the former with 4 to 6 digits of accuracy in all cases. In Fig. \ref{fig:IsingCritExp} we plot the scaling dimensions of magnitude $\Delta\le 2.5$ obtained from the scale-invariant MERA, which correspond to the primary fields and their descendants, organized both according to the locality $g$ and the parity $p$ of the corresponding scaling operators. Local scaling operators ($g=+1$) with even parity ($p=+1$) form the two conformal towers of the primary fields identity $\mathbb{I}$ and energy $\epsilon$ of the Ising CFT, whereas those with odd parity ($p=-1$) form the conformal tower of the primary field spin $\sigma$. Non-local scaling operators ($g=-1$) with even parity ($p=+1$) form the conformal tower of the disorder operator $\mu$, and those with odd parity ($p=-1$) are organized according to two towers corresponding to the fermion operators $\psi$ and $\bar{\psi}$. The numerical results from the scale-invariant MERA are seen to accurately reproduce the smallest scaling dimensions, those with $\Delta\le 2.5$, from all six conformal towers of the Ising CFT \cite{Francesco97,Henkel99}. 

\begin{table}[!htbp]
\centering
\caption[]{Scaling dimensions of the primary fields of the Ising CFT.}
\renewcommand{\arraystretch}{1.2}
\setlength\tabcolsep{5pt}
\begin{tabular}{ |l | l | l |}
\hline
$\Delta^{\mbox{\tiny exact}}$  & $~\Delta^{\mbox{\tiny MERA}}_{\mbox{\tiny $\chi=32$}}~$ & Error  \\
\hline
$\Delta_\sigma$=0.125         &  $~$ 0.1249998         & $2\times 10^{-4}$ $\%$ \\
  $\Delta_\epsilon$=1           &  $~$ 1.0001139         & 0.011       $\%$ \\
  $\Delta_\mu$=0.125            &  $~$ 0.1250002         & $2\times 10^{-4}$ $\%$ \\
  $\Delta_\psi$=0.5             &  $~$ 0.4999959         & $8\times 10^{-4}$ $\%$ \\
  $\Delta_{\bar\psi}$=0.5       &  $~$ 0.4999963         & $7\times 10^{-4}$ $\%$ \\
  \hline\end{tabular}
\label{tab:IsingExp}       
\end{table}

We have also computed the OPE coefficients $C_{\alpha\beta\gamma}$ for all primary fields, obtained by analyzing three-point correlators as described in Refs. \cite{Pfeifer09}. Table \ref{tab:IsingOPE} shows the numerical estimate of all non-vanishing OPE coefficients. Once again the results match their exact values to within several digits of accuracy.
\begin{table}[!tbp]
\centering
\caption[]{OPE coefficients for the local and non-local primary fields of the Ising CFT.}
\renewcommand{\arraystretch}{1.2}
\setlength\tabcolsep{5pt}
\begin{tabular}{ |l | l | l |}
\hline
$~~~~~C^{\mbox{\tiny exact}}$  & $~C^{\mbox{\tiny MERA}}_{\mbox{\tiny $\chi=32$}}~$ & error  \\
\hline
$C_{\epsilon, \sigma, \sigma} = 1/2$  &  $~$ 0.50008  & 0.016$\%$ \\
   $C_{\epsilon, \mu, \mu} = -1/2$  &  $~$ -0.49997  & 0.006$\%$    \\

   $C_{\psi,     \mu,    \sigma} = \frac{e^{-i\pi/4}}{\sqrt{2}}$     &  $\frac{1.00068e^{-i\pi/4}}{\sqrt{2}}$ $~$ & 0.068$\%$ \\ 
   $C_{\bar\psi,     \mu,    \sigma} = \frac{e^{i\pi/4}}{\sqrt{2}}$  &  $\frac{1.00068e^{i\pi/4}}{\sqrt{2}}$ $~$  & 0.068$\%$   \\

   $C_{\epsilon, \psi, \bar\psi} = i$     &  $1.0001 i$  $~$ & 0.010$\%$ \\
   $C_{\epsilon, \bar\psi, \psi} = -i$    &  $-1.0001 i$ $~$ & 0.010$\%$   \\ \hline\end{tabular}
\label{tab:IsingOPE}       
\end{table}
Thus, not only have we been able to identify the \emph{entire} field content $\{\mathbb{I},\epsilon,\sigma, \psi,\bar{\psi},\mu\}$ of the Ising CFT from a simple and rather inexpensive analysis of a quantum spin chain, but we can now also identify all possible subsets of primary fields that close a subalgebra by inspecting Table \ref{tab:IsingOPE}. Indeed, it follows that we have the fusion rules
\begin{eqnarray}
 \epsilon \times \epsilon = \mathbb{I}, ~~~\sigma \times \sigma = \mathbb{I} + \epsilon,~~~ \sigma \times \epsilon = \sigma, \\
\mu \times \mu = \mathbb{I} + \epsilon, ~~~\mu \times \epsilon = \mu, \\
\psi \times \psi = \mathbb{I},~~~\bar{\psi} \times \bar{\psi} = \mathbb{I},\\
\psi \times \bar{\psi} = \epsilon,~~~
\psi \times \epsilon = \bar{\psi}, ~~~\bar{\psi} \times \epsilon = \psi, \label{eq:sCe25}
\end{eqnarray}
(as well as other, such as $\sigma \times \mu = \psi + \bar{\psi}$, etc) from where we see that $\{\mathbb{I},\epsilon\}$ and $\{\mathbb{I},\epsilon,\sigma\}$ close subalgebras of local primary fields, whereas $\{\mathbb{I},\epsilon,\mu\}$ and $\{\mathbb{I},\epsilon,\psi,\bar{\psi}\}$ close subalgebras that contain both local and non-local primary fields, where locality is relative to the spin variables.

\section{Conclusions} \label{sect:Conclusions}

In this manuscript we have presented an introduction to the scale-invariant MERA and its application to the study of quantum critical systems. The main strength of MERA, when applied to quantum critical systems, is that it can explicitly incorporate scale invariance. This facilitates enormously the computation of scaling dimensions (equivalently, critical exponents) and of other properties that characterize a quantum phase transition.

Direct comparison with an MPS shows that, while the later is more efficient at computing local observables such as the ground state energy, the MERA produces significantly more accurate correlators at long distances. In addition, from the MERA it is straightforward to identify the scaling operators of the theory, as well as their scaling dimensions and operator product expansion, producing accurate conformal data that can be used to unambiguously identify the underlying CFT.

Here we have considered homogeneous systems. However, the scale-invariant MERA has been successfully generalized to critical systems where translation invariance is explicitly broken, as it is the case of a critical system with a boundary, with an impurity, or the interface between two critical systems \cite{Evenbly10e,Silvi10,Evenbly11b}. In all these scenarios translation invariance is no longer present, but exploitation of scale invariance still produces a MERA algorithm with a cost $O(1)$ (that is, independent of the lattice size $N$), and therefore infinite systems can be addressed. Again, scaling operators associated to boundaries, defects and interfaces can be easily extracted from the simulations.

Finally, much of the MERA formalism for critical systems in $D=1$ dimensions is also directly applicable to $D=2$ dimensions, including the characterization of scaling dimensions. However, due to significantly larger computational costs, so far only small values of $\chi$ have been used in actual computations. Thus, further progress needs to be made in reducing computational costs before the scale-invariant MERA becomes a viable approach to quantum criticality also in $D=2$ dimensions.

\appendix

\section{Optimization algorithm for the scale-invariant MERA} \label{sect:ScaleImp}

In this section we describe an algorithm to optimize the scale-invariant MERA to approximate the ground state of a critical system. It is based on modifying the optimization algorithm for a regular MERA of Ref. \cite{Evenbly09} and has been previously sketched in Refs. \cite{Evenbly09,Pfeifer09}, although the present implementation differs in some details. 
 
A scale-invariant MERA is composed of a small number $M$ of transitional layers $\left\{ U^{[0]}, U^{[1]}, \ldots\right.$ $\left. , U^{[M-1]} \right\}$ followed by an infinite sequence of (identical) scale-invariant layers, here denoted with an asterisk, $U^{*}$. We shall henceforth use similar asterisk notation for all operators and tensors associated to the scale-invariant layers. Each of the transitional layers $U^{[\tau]}$ are characterized by a single isometry $w^{[\tau]}$ and disentangler $u^{[\tau]}$; similarly, the scale-invariant layers $U^{*}$ are characterized by a single isometry $w^{*}$ and disentangler $u^{*}$. The goal of this section is to describe how these tensors can be optimized in such a way that, given a critical Hamiltonian, the scale-invariant best approximates its ground state.

We begin, in Sect. \ref{sect:AlgBuildBlock}, by describing the key building blocks of the optimization algorithm, which are (i) coarse-graining of the Hamiltonian; (ii) fine-graining of the density matrix; and (iii) optimization of one tensor of the MERA. Then, in Sect. \ref{sect:OptAlg}, we describe how these algorithmic building blocks can be put together to form an iterative optimization scheme. Finally, Sect. \ref{sect:CompTricks} describes computational tricks to improve convergence and accuracy.

\begin{figure}[!tbph]
\begin{center}
\includegraphics[width=14cm]{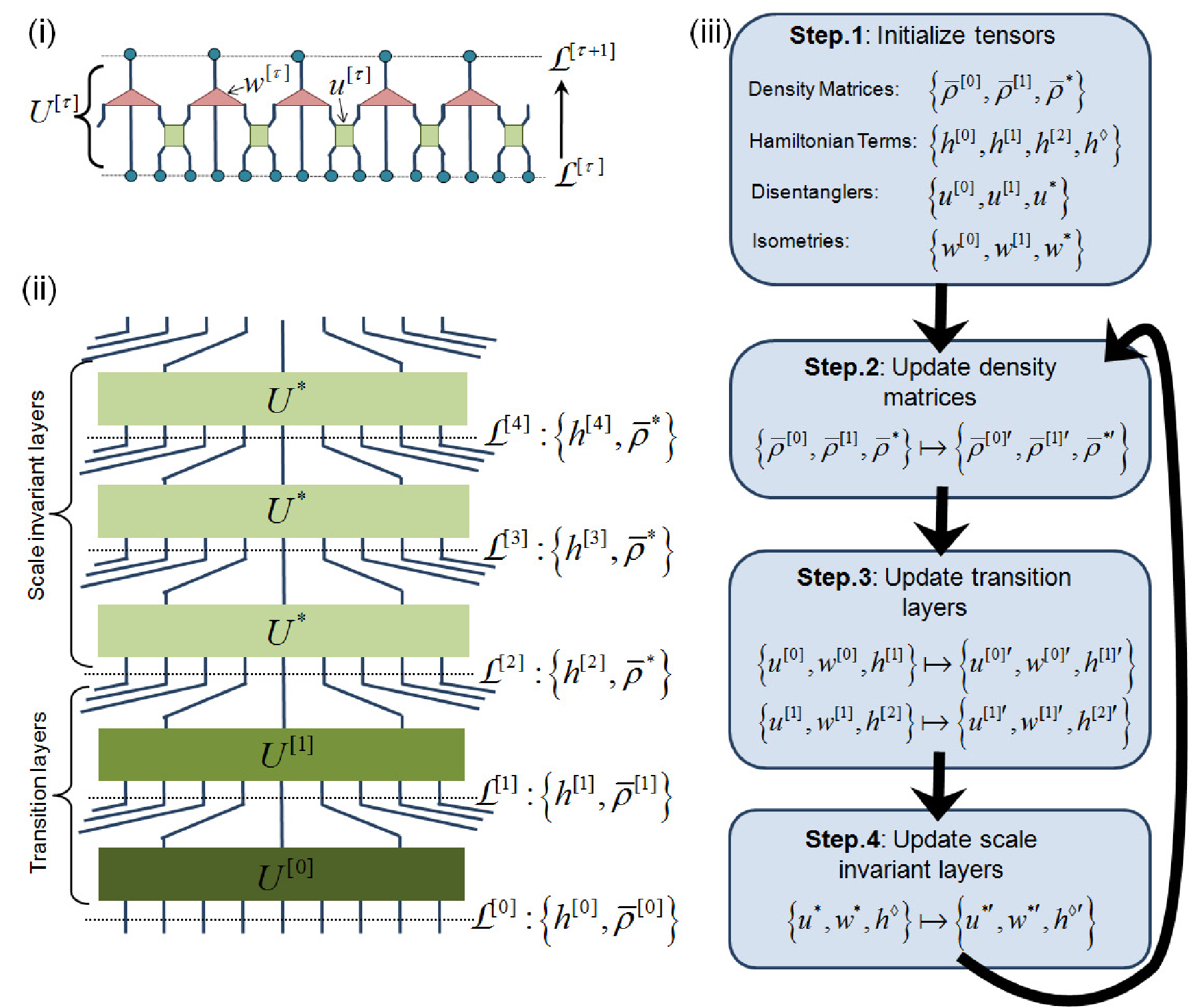}
\caption{(i) Each layer $U^{[\tau]}$ of the MERA may be thought of as a coarse-graining transform between an initial lattice $\mathcal{L}^{[\tau]}$ and a coarser lattice $\mathcal{L}^{[\tau+1]}$. (ii) A schematic representation of the scale-invariant MERA, here with two transitional layers $U^{[0]}$ and $U^{[1]}$ followed by an infinite number of identical scale-invariant layers $U^{*}$. (iii) The steps for optimizing a scale-invariant MERA as described Sect. \ref{sect:OptAlg}. }
\label{fig:AlgDef}
\end{center}
\end{figure}

\subsection{Building blocks} \label{sect:AlgBuildBlock}

\subsubsection{Coarse-graining of the Hamiltonian}\label{sect:LiftHam}

A key part of the MERA optimization algorithm is the coarse-graining of the Hamiltonian. Given an initial Hamiltonian $H^{[0]}=\sum\nolimits_r h^{[0]} (r,r+1)$ that decomposes as a sum of identical local terms $h^{[0]}(r,r+1) = h^{0}$, we wish to construct, for any level $\tau$, the coarse-grained Hamiltonian $H^{[\tau]}=\sum\nolimits_r h^{[\tau]} (r,r+1)$, defined as
\begin{equation}
H^{[\tau]}  \equiv \left( U^{[\tau-1]} \right)^\dag  H^{[\tau-1]} \left( U^{[\tau-1]} \right). \label{eq:a1e1}
\end{equation}
The coarse-graining of an operator that decomposes as a sum of local operators, such as a local Hamiltonian, can be achieved with the ascending superoperator formalism introduced in Sect. \ref{sect:ERfoundation} for the coarse-graining of a single local operator. The coarse-grained Hamiltonian coupling $h^{[\tau  + 1]}$ is obtained by enacting the (left,center,right) ascending superoperators ${\mathcal A}_L, {\mathcal A}_C$ and ${\mathcal A}_R$ on the Hamiltonian term $h^{[\tau]}$, 
\begin{align}
 h^{[\tau+1]}  &\equiv \mathcal A_L^{[\tau ]} \left( h^{[\tau ]}  \right) + \mathcal A_C^{[\tau ]} \left( h^{[\tau ]}  \right) + \mathcal A_R^{[\tau ]} \left( {h^{[\tau ]} } \right) \nonumber \\ 
  &= 3 \mathcal{\bar A}^{[\tau ]} \left( {h^{[\tau ]} } \right) \label{eq:a1e2}
\end{align}
where $\mathcal{\bar A} $ is the average of the three ascending superoperators. The diagrammatic representation of the tensor network described by Eq. \ref{eq:a1e2} is shown in Fig. \ref{fig:MERAmanip}(i).

\subsubsection{Fine-graining of the density matrix} \label{sect:LowDensity}
\begin{figure*}[!p]
\begin{center}
\includegraphics[width=12cm]{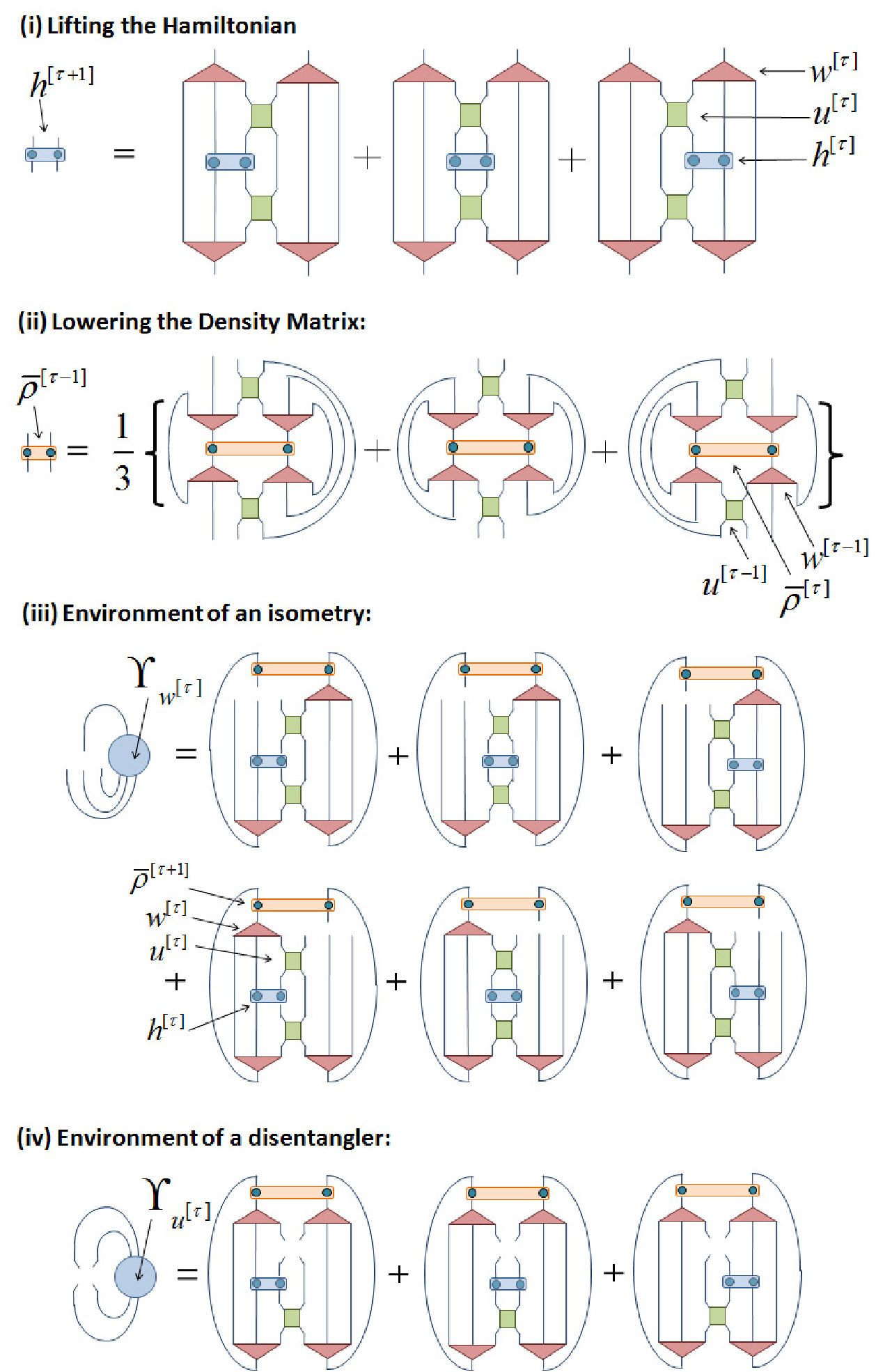}
\caption{This figure displays the full set of tensor network contractions required to optimize a ternary MERA. (i) The tensor network contractions required to coarse-grain a local Hamiltonian, see Sect. \ref{sect:LiftHam}. (ii) The tensor network contractions required to fine-grain the average two-site density matrix, see Sect. \ref{sect:LowDensity}. (iii) The linearized environment $\Upsilon_w$ of an isometry $w$ and (iv) the linearized environment $\Upsilon_u$ of a disentangler $u$, see Sect. \ref{sect:LinOpt}.}
\label{fig:MERAmanip}
\end{center}
\end{figure*}

Another key part of the MERA optimization algorithm is fine-graining the density matrix. Given the average two-site density matrix $\bar \rho^{[\tau]}$ defined on lattice $\mathcal{L}^{[\tau]}$, we wish to construct the average two-site density matrix $\bar \rho^{[\tau -1]}$ on lattice $\mathcal{L}^{[\tau -1]}$. This is accomplished through the average descending superoperator $\mathcal{\bar D}$ introduced in Sect. \ref{sect:LocDensity}, which acts as
\begin{equation}
\bar \rho ^{[\tau-1]} = \mathcal{\bar D} ^{[\tau-1]} \left( \bar \rho^{[\tau]} \right). \label{eq:a1e3}
\end{equation}
From iteration of Eq. \ref{eq:a1e3} one can construct the average two-site density matrices $\bar \rho ^{[\tau ']}$ for all $\tau ' < \tau$. The diagrammatic representation of the tensor network described by Eq. \ref{eq:a1e3} is shown in Fig. \ref{fig:MERAmanip}(ii).

\subsubsection{Optimization of one tensor of the MERA} \label{sect:LinOpt}

In order to approximate the ground state of a Hamiltonian $H$, the disentanglers $u$ and isometries $w$ that define a scale-invariant MERA $\ket{\Psi}$ should be chosen to minimize the energy $E \equiv \bra{\Psi} H \ket{\Psi}$. We shall proceed by updating one tensor of the MERA at a time, while holding the rest of the tensors fixed. Given a tensor to be updated, we note that the energy $E$ depends quadratically on that tensor and its conjugate. Here we employ the linearized optimization scheme of Ref. \cite{Evenbly09}, which we sketch next (a justification for linearizing the cost function and further details can be found in Ref. \cite{Evenbly09}).

The update of an isometry $w$ relies on computing its \textit{environment} $\Upsilon_{w}$, which represents a factorization of $E=\bra{\Psi} H \ket{\Psi}$ to isolate its dependence on $w$,
\begin{equation}
	E = \tr(w \Upsilon_{w}) + k_1, \label{eq:a1e4}
\end{equation}
with $k_1$ an irrelevant constant. [$\Upsilon_{w}$ is the derivative of $\bra{\Psi} H \ket{\Psi}$ with respect to $w$ while keeping $w^{\dagger}$ fixed].  The updated isometry $w'$ that minimizes the linearized energy ---that is, such that $\tr(w^{[\tau]'} \Upsilon_{w^{[\tau]}})$ is minimal--- is given by $w ' = -V_2 V_1^{\dagger}$, where $V_1$ and $V_2$ are obtained from the singular value decomposition (SVD) of the environment, $\Upsilon_{w} = V_1 S V_2^{\dagger}$. We then proceed by replacing that particular isometry $w$ with $w'$ in the MERA, which now represents a new state $\ket{\Psi'}$. We emphasize that the energy $E' = \bra{\Psi'}H\ket{\Psi'}$ is computed by replacing both $w$ and $w^\dagger$ with $w'$ and $w'^{\dagger}$, and it is therefore not given by $\tr(w' \Upsilon_{w}) + k_1$. In other words, $w'$ does not minimize the expectation value of the Hamiltonian (in fact, the energy could even rise!). However, the approach works well in practice. The update of a disentangler $u$ follows an analogous procedure, involving the SVD of the corresponding environment $\Upsilon_{u}$.

The environments of an isometry $w^{[\tau]}$ and disentangler $u^{[\tau]}$ at layer $\tau$ depend only on a small number of other tensors,  
\begin{eqnarray}
\Upsilon_{w^{[\tau]}} &=& \Upsilon_{w^{[\tau]}} \left( u^{[\tau]},w^{[\tau]},\bar \rho ^{[\tau+1]}, h^{[\tau]} \right), \label{eq:a1e5a}\\
\Upsilon_{u^{[\tau]}} &=& \Upsilon_{u^{[\tau]}} \left( u^{[\tau]},w^{[\tau]},\bar \rho ^{[\tau+1]}, h^{[\tau]} \right), \label{eq:a1e5}
\end{eqnarray}
as depicted in Figs. \ref{fig:MERAmanip}(iii)-(iv).

\subsection{Optimization algorithm} \label{sect:OptAlg}

For concreteness, we describe the optimization of a scale-invariant MERA with $M=2$ transitional layers as depicted Fig. \ref{fig:AlgDef}(ii). The ansatz is then completely characterized by three isometries $\left\{ w^{[0]},  w^{[1]},  w^{*} \right\} $ and three disentanglers $\left\{ u^{[0]},  u^{[1]},  u^{*} \right\} $, where the isometry $w^{*}$ and the disentangler $u^{*}$ define the scale-invariant layers. For purposes of the optimization algorithm, one should also store in computer memory the two-site Hamiltonian couplings $\left\{h^{[0]}, h^{[1]}, h^{[2]}, h^{\diamondsuit}\right\}$. Here $h^{\diamondsuit}$ is a scale-averaged Hamiltonian, to be introduced later in Eq. \ref{eq:a1e10}. We also store the average two-site density matrices $\left\{ \bar \rho^{[1]}, \bar{\rho}^{*}\right\}$, where we recall that $\bar \rho^{*}$ is the two-site density matrix any of the scale-invariant layers. We now proceed to describe the optimization algorithm for scale-invariant MERA, as outlined Fig. \ref{fig:AlgDef}(iii). In an iteration of the algorithm, comprising Sect. \ref{sect:UpdateDensity}, Sect. \ref{sect:UpdateTransition} and Sect. \ref{sect:UpdateScale}, each of the stored tensors is updated to a new version of itself denoted with a prime, e.g. $u^{[0]}\mapsto u^{[0]}{'}$.

\subsubsection{Initialization} \label{sect:OptInit}
The preliminary step of the algorithm is to initialize the tensors which define the MERA. It is most often sufficient to initialize the tensors to be random, though in certain situations, for instance if one has prior knowledge about the ground state of the Hamiltonian, it is useful to use that information. Random isometries $\left\{ w^{[0]}, w^{[1]}, w^{*}\right\}$ can be obtained through singular value decomposition of an appropriately sized random rectangular matrix. The disentanglers $\left\{ u^{[0]}, u^{[1]}, u^{*}\right\}$ can also be initialized randomly, or as the identity operator. Notice that the local Hamiltonian term $h^{[0]}$ is given as the input of the algorithm (it describes the critical Hamiltonian whose ground state is to be found), whereas the Hamiltonian terms $\left\{ h^{[1]}, h^{[2]}, h^{\diamondsuit }\right\}$ and density matrices $\left\{\bar \rho^{[1]},\bar \rho^{*}\right\}$ will be generated during the first iteration of the algorithm.

\subsubsection{Updating the density matrices} \label{sect:UpdateDensity}

The first step of each iteration is to update the two-site reduced density matrices. As explained Sect. \ref{sect:LocDensity}, the scaling density matrix $\bar \rho ^{*}{'}$ is defined as the dominant eigenoperator of the average descending superoperator $\mathcal{\bar D}^{*}$ associated to the scale-invariant layers,
\begin{equation}
\bar \rho^{*}{'} = \mathcal{\bar D}^{{*}} \left( \bar{\rho}^{*}{'} \right). \label{eq:a1e6}
\end{equation}
One can solve Eq. \ref{eq:a1e6} for $\bar \rho^{*}{'}$ with a sparse diagonalization technique such as the Lanczos method, where the scaling density matrix $\bar \rho ^{*}$ from the previous iteration (if any) can be reused as the starting point for the Lanczos method to accelerate convergence. The updated density matrix $\bar \rho ^{[1]}{'}$ is then obtained by fine-graining $\bar{\rho}^{*}{'}$ with the descending superoperator $\mathcal{\bar D}^{[1]}$,
\begin{equation}
\bar \rho ^{[1]}{'} = \mathcal{\bar D}^{[1]} \left( \bar \rho ^{*}{'} \right). \label{eq:a1e7}
\end{equation}

\subsubsection{Updating the transitional layers} \label{sect:UpdateTransition}
The next step of the iteration is to update the transitional layers of the MERA. The tensors $w^{[0]}$ and $u^{[0]}$, which comprise the first transitional layer $U^{[0]}$, are updated by first computing their linearized environments,
\begin{align}
\Upsilon_{w^{[0]}} &= \Upsilon_{w^{[0]}} \left( u^{[0]},w^{[0]},\bar \rho ^{[1]}{'}, h^{[0]} \right), \nonumber \\ 
\Upsilon_{u^{[0]}} &= \Upsilon_{u^{[0]}} \left( u^{[0]},w^{[0]},\bar \rho ^{[1]}{'}, h^{[0]} \right) , \label{eq:a1e8}
\end{align}
from which the updated tensors $w ^{[0]}{'}$ and $u ^{[0]}{'}$ are obtained through SVD, see Sect. \ref{sect:LinOpt}. Together the updated isometry $w ^{[0]}{'}$ and disentangler $u ^{[0]}{'}$ define the ascending superoperator $\mathcal{\bar A} ^{[0]}$, which is used to coarse-grain the Hamiltonian $h^{[0]}$,
\begin{equation}
h ^{[1]}{'} = 3 \mathcal{\bar A} ^{[0]} \left( h^{[0]} \right), \label{eq:a1e9}
\end{equation}
and obtain the updated Hamiltonian coupling $h ^{[1]}{'}$, as described Sect. \ref{sect:LiftHam}. The second transitional layer $U^{[1]}$ is then updated, in the same manner as described here for the first transitional layer, to obtain updated tensors $\left\{ w ^{[1]}{'}, u ^{[1]}{'}, h ^{[2]}{'} \right\}$.

\subsubsection{Updating the scale-invariant layers} \label{sect:UpdateScale}

The final step of the iteration is to update the isometry $w^{*}$ and disentangler $u^{*}$ associated to the scale-invariant layers of the MERA. Updating the scale-invariant layers is more complicated than updating the transitional layers since the scaling tensors $w ^{*}$ and $u ^{*}$ do not appear in just one layer but in all layers $U^{[\tau]}$ for $\tau\ge M$, with $M$ the number of transitional layers). To update these tensors, we construct linearized environments that average the contributions coming from all length scales $\tau \ge M$. Computation of these scale-averaged environments is simplified \cite{Evenbly09} by the introduction of the scale-averaged Hamiltonian term $h ^{\diamondsuit}$, which for the case of $M$ transitional layers is defined as
\begin{equation}
h^{\diamondsuit}  \equiv \sum\limits_{\tau  = M}^\infty  {\frac{1}{{3^{\tau  - M} }}h^{[\tau ]} }  = h^{[M]}  + \frac{1}{3}h^{[M+1]}  + \frac{1}{9}h^{[M+2]}  +  \ldots, \label{eq:a1e10}
\end{equation}
where the factor of $(1/3)^{\tau - M}$ arises from the fact that layer $U^{[\tau]}$ contains three times as many tensors as layer $U^{[\tau+1]}$ \cite{Evenbly09}. 

Note that for a critical Hamiltonian (with negligible or sufficiently suppressed RG-irrelevant terms) and a scale invariant MERA that has already been optimized, scale-invariance implicitly assumes that $h^{[\tau+1]}\propto h^{[\tau]}$ for all $\tau\geq M$, and therefore the scale-averaged Hamiltonian $h^{\diamondsuit}$ is simply proportional to $h^{[M]}$. However this property will only hold for a MERA which has been properly optimized; \emph{during} the optimization procedure it is necessary to explicitly average over all scale-invariant layers when constructing environments. Computing the scale-averaged Hamiltonian $h ^{\diamondsuit}$ directly through Eq. \ref{eq:a1e10} is not feasible due to the infinite summation. One strategy to obtain an approximation to $h ^{\diamondsuit}$ is to use a partial summation of Eq. \ref{eq:a1e10} which stops at some $\tau=T$. In practice, however, a useful estimate of the updated scale-averaged Hamiltonian $h ^{\diamondsuit}{'}$ is already obtained from the operator $h^{\diamondsuit }$ from the previous iteration through
\begin{equation}
h^{\diamondsuit}{'} \approx h^{[M]}{'}  + \bar{\cal{A}}^{*}\left( {h^{\diamondsuit } } \right). \label{eq:a1e11}
\end{equation}
This estimate of $h^{\diamondsuit}{'}$ is accurate provided that the effective Hamiltonians are only changing by small amounts between iterations, i.e. $h^{[\tau]}{'} \approx h^{[\tau]}$ for all $\tau$, which becomes a better approximation as the optimization nears convergence. Once the new $h^{\diamondsuit}{'}$ has been computed then the environments of the scaling isometry $w^{*}$ and disentangler $u^{*}$ are computed,
\begin{align}
\Upsilon_{w^{*}} &= \Upsilon_{w^{*}} \left( u^{*},w^{*},\bar {\rho}^{*}{'}, h^{\diamondsuit}{'} \right), \nonumber \\ 
\Upsilon_{u^{*}} &= \Upsilon_{u^{*}} \left( u^{*},w^{*},\bar{\rho}^{*}{'}, h^{\diamondsuit}{'} \right), \label{eq:a1e12}
\end{align}
from which, as usual, one obtains the updated tensors $w^{*}{'}$ and $u^{*}{'}$ by SVD. This concludes a single iteration of the optimization algorithm for scale-invariant MERA; the next iteration can begin again at Sect. \ref{sect:UpdateDensity}.

\subsection{Computational tricks} \label{sect:CompTricks}

\begin{figure}[!tbhp]
\begin{center}
\includegraphics[width=12cm]{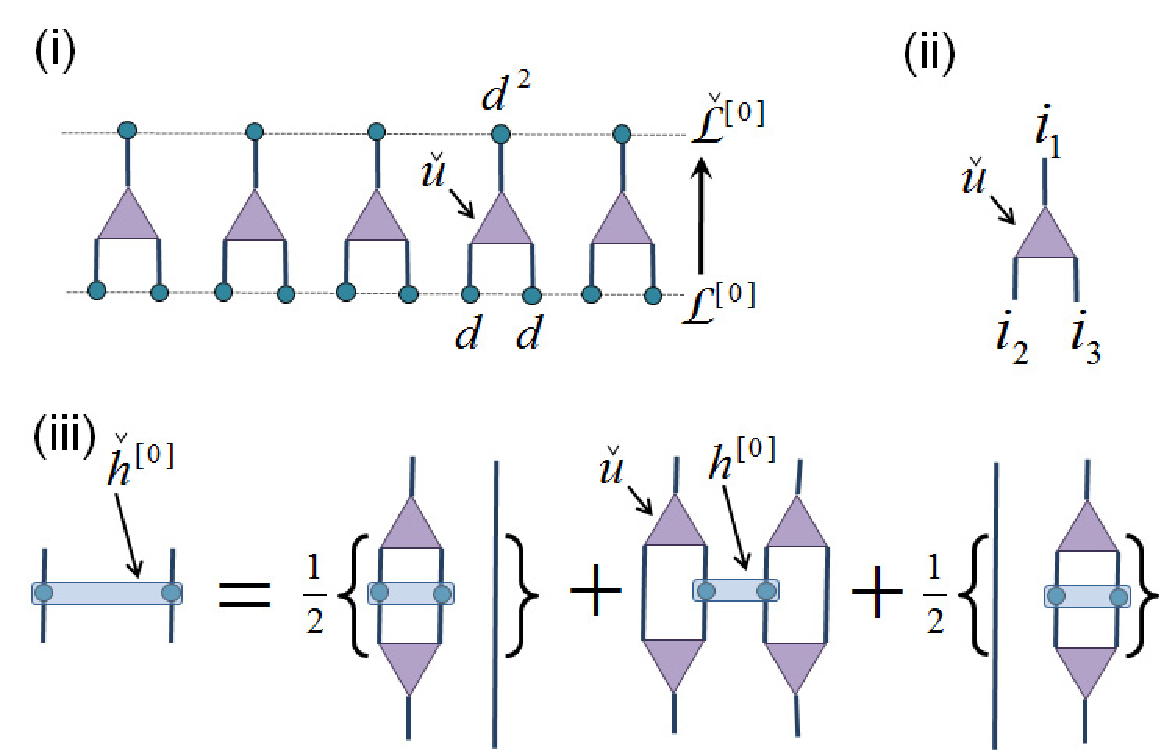}
\caption{(i) The preliminary blocking groups two $d$-dimensional sites of the original lattice ${\cal{L}}^{[0]}$ into a single $d^2$-dimensional site of the lattice ${\cal{\check L}}^{[0]}$. (ii) The tensor $\check u_{i_2 ,i_3 }^{i_1 }$ is equivalent to the identity, see Eq. \ref{eq:a1e15}. (iii) Under the preliminary blocking the initial Hamiltonian $H^{[0]}=\sum_r h^{[0]} {(r,r+1)}$ on lattice ${\cal{L}}^{[0]}$ is mapped to an equivalent Hamiltonian $\check H^{[0]}=\sum_r \check h^{[0]} {(r,r+1)}$ on lattice ${\cal{\check L}}^{[0]}$.}
\label{fig:PrelimCG}
\end{center}
\end{figure}

\subsubsection{Preliminary blocking} \label{sect:PrelimCG}

Given a local Hamiltonian defined on a lattice $\mathcal{L}^{[0]}$, where each lattice site is described by a Hilbert space $\mathbb{V}$ of finite dimension $d$, in certain cases it may be convenient to perform a preliminary blocking of e.g. two sites of $\mathcal{L}^{[0]}$ into a single site of dimension $\check d =d^2$ in the coarser lattice ${\cal{\check L}}^{[0]}$. This preliminary blocking, which maps the initial Hamiltonian $H^{[0]}=\sum h^{[0]}$ on lattice $\mathcal L^{[0]}$ to a new Hamiltonian $\check H^{[0]} =\sum \check h^{[0]}$ defined on a coarser lattice ${\cal{\check L}}^{[0]}$, simply amounts to re-expressing the Hamiltonian in a different form, i.e. $H^{[0]}$ and $\check H^{[0]}$ are different representations of the same Hamiltonian. An example how the Hamiltonian can be redefined through a preliminary blocking, for the specific case of blocking two sites into one, is depicted Fig. \ref{fig:PrelimCG}, with $\check u$ a tensor equivalent to the identity,
\begin{equation}
\check u_{i_2 ,i_3 }^{i_1 }  = \left\{ {\begin{array}{*{20}l}
   {1,\; \; i_1  = i_2 + d \: i_3 }  \\
   {0,\; \; {\rm{else}}}  \\
\end{array}} \right. \label{eq:a1e15}
\end{equation}
where indices $i_2$ and $i_3$ take $d$ values each whereas index $i_1$ takes $d^2$ values. This preliminary blocking, which does not increase (to leading order) the contraction cost of the MERA provided $d^2$ is less than the bond dimension $\chi$ used in higher layers of the MERA, has two potential advantages. Firstly, it can reduce a next-nearest neighbor Hamiltonian $H^{[0]}$ into a nearest neighbor Hamiltonian $\check H^{[0]} $ (of larger local dimension), which can then be easily treated with the ternary MERA. The second advantage is that it transforms a state that would otherwise be translation-invariant by shifts of two sites into a state that is translation-invariant by shifts of just one site, which can then be represented directly with a translation-invariant ternary MERA.

\subsubsection{Shifting the Hamiltonian spectrum} \label{sect:HamShift}

Let us assume we are interested in obtaining the ground state MERA of a given local Hamiltonian $H^{[0]}=\sum_r h^{[0]} (r,r+1)$. Prior to the optimization it can be useful to shift the spectrum of the Hamiltonian by adding or subtracting contributions of the identity,
\begin{equation}
h^{[0 ]}  \mapsto {h}_\alpha^{[0]} \equiv h^{[0]}  - \alpha \; \mathbb I, \label{eq:a1e16}
\end{equation}
where $\mathbb I$ is the identity operator on two sites, such that the shifted Hamiltonian $H_\alpha^{[0]}=\sum\nolimits h^{[0]}_\alpha$ is negative defined. This can be achieved by choosing $\alpha$ in Eq. \ref{eq:a1e16} as the largest eigenvalue of $h^{[0]}$. Having a negative defined Hamiltonian ensures that the optimization targets the low-energy, i.e. ground state, subspace (the optimization algorithm, based upon extremizing the energy of the state, could otherwise target the high-energy subspace). 


It is also useful to similarly shift the spectrum of the coarse-grained Hamiltonian that arise during the optimization, by replacing $h^{[\tau]} \mapsto h_\alpha^{[\tau]}$ where $\alpha$ as the largest eigenvalue of $h^{[\tau]}$, since this is seen to speed-up convergence during the optimization of the MERA.

\subsubsection{Converging the number of transitional layers} \label{sect:TransLay}

In order to obtain an accurate approximation to the ground state of critical Hamiltonian $H$ with a scale invariant MERA it is important to use enough transitional layers to sufficiently suppress the effect of any RG irrelevant terms present in $H$, which can break scale-invariance at short length scales. For a given critical $H$ the appropriate number transitional layers is not known \emph{a priori}, and must be found by trial and error. 
We proceed in the following way. Suppose that we have already optimized a MERA with $M=2$ transitional layers, that is, characterized by $\left\{ {U^{[0]} ,U^{[1]} ,U^{*} } \right\}$. We can then add an additional transitional layer, $M=3$, so that now the MERA is given by 
$\left\{ {U^{[0]} ,U^{[1]} ,U^{[2]} ,U^{*}} \right\}$, where we initially set $U^{[2]} =U^{*}$, and re-optimize the tensors as to best approximate the ground state of Hamiltonian $H$. We then keep adding more transitional layers, and re-optimizing the ansatz, until addition of an extra transitional layer does not produce significantly different results under re-optimization in terms e.g. of the ground energy, scaling dimensions, etc.

\section{Comparison of MERA schemes} \label{sect:SchemeCompare}

In Sect. \ref{sect:ChoiceScheme} three different implementations of the MERA in $D=1$ dimensions were described: the binary MERA, the ternary MERA and the modified binary MERA. The three schemes differ in how the computational cost (incurred e.g. in optimizing the ansatz and in computing the expectation value of local observables) scales with the bond dimension $\chi$, namely as $O(\chi^9)$, $O(\chi^8)$ and $O(\chi^7)$, respectively. In addition, the three schemes also differ in the strength of disentangling, which means that for the same value of $\chi$ the numerical accuracy of the results will vary across the three schemes. In this section we investigate which of the schemes provides the most accurate ground state energy for a fixed computation cost.

\begin{figure}[!tbph]
\begin{center}
\includegraphics[width=10cm]{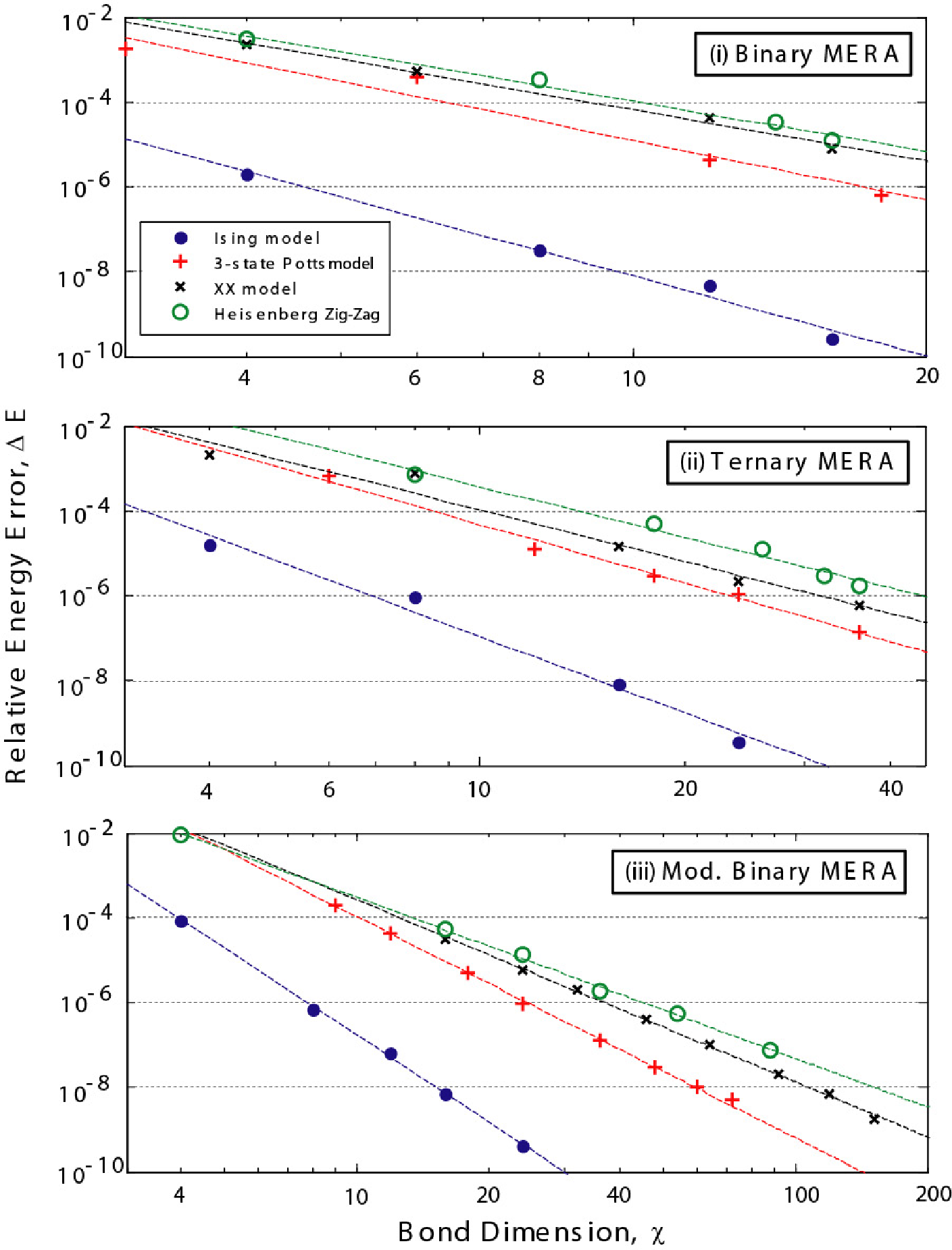}
\caption{Relative energy error $\Delta E$ in the ground state of critical spin chains, as a function of the MERA bond dimension $\chi$, comparing three different MERA schemes.}
\label{fig:SchemeEnergy}
\end{center}
\end{figure}

For this purpose, we have optimized each of the three MERA schemes so as to approximate the ground states of the critical Hamiltonians defined Eqs. \ref{eq:sBe1}-\ref{eq:sBe4} in Sect. \ref{sect:MERABench}. Fig. \ref{fig:SchemeEnergy} shows the relative energy error $\Delta E$, defined as
\begin{equation}
\Delta E \equiv (E_{\textrm{exact}} - E_{\textrm{numeric}})/ E_{\textrm{exact}},\label{eq:a2e0}
\end{equation}
as a function of bond dimension $\chi$. For all three schemes the relative energy error $\Delta E$ is seen to scale as a power of the bond dimension $\chi$, 
\begin{equation}
\Delta E\approx a \chi^{-b}, \label{eq:a2e1}
\end{equation}
for some coefficients `$a$' and `$b$' that depend both on the MERA scheme in use and the spin model under consideration. The sets of these coefficients, obtained from linear fits in Fig. \ref{fig:SchemeEnergy}, are displayed in Table \ref{tab:SchemeEnergy}. For each of the four spin models separately, the data shows that while the coefficient `$a$' depends considerably on the MERA scheme (with variations by one or two orders of magnitude), the coefficient `$b$' is remarkably constant across the schemes (with variations of about $15\%$). These results indicate that, while using a scheme with more powerful disentangling (such as the binary MERA scheme) can reduce the energy error by a considerable, fixed multiplicative factor (as expressed by a significantly smaller constant `$a$'), the power `$b$' of $\chi$ for a given critical model is not significantly affected by the disentangling power of the scheme, even though the latter does affect significantly how the computational cost scales with $\chi$. This suggests that the modified binary MERA scheme, which has the smallest disentangling power, will be the most cost effective at large $\chi$, since its cost scales as a smaller power of $\chi$ and the accuracy scales roughly as the same power of $\chi$ as the other schemes.

\begin{figure}[!tbhp]
\begin{center}
\includegraphics[width=10cm]{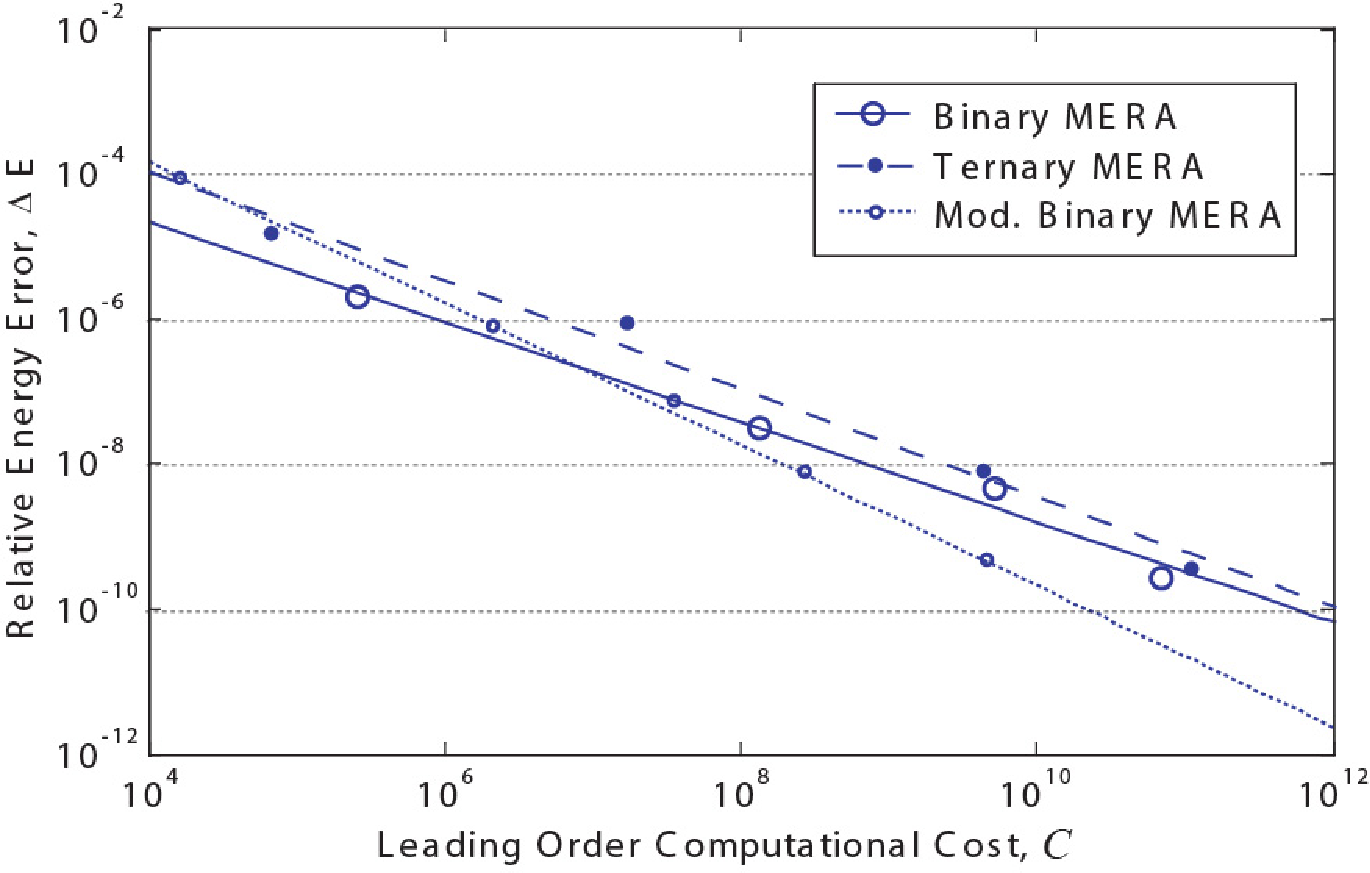}
\caption{Relative energy error scaling in ground state MERA calculations, for the three different MERA schemes as described in Sect. \ref{sect:ChoiceScheme}, in the quantum critical Ising model plotted as a function of the leading order computational cost $C$ for the optimization.}
\label{fig:SchemeCompare}
\end{center}
\end{figure}

Fig. \ref{fig:SchemeCompare} shows $\Delta E$ directly as a function of the computational cost for the Ising model. Similar results are obtained for the other models. The costs have been approximated to be $C=\chi^9$ for binary, $C=\chi^8$ for ternary and $C=\chi^7$ for modified binary MERA, where $C$ is measured in floating-point operations. The exact expression for the cost would include a multiplicative constant of order one that does not change significantly across the schemes and has been ignored for simplicity. The plot confirms that, for large $\chi$, the modified binary MERA scheme, with cost $C=\chi^7$, is the scheme that provides the most accurate results for a fixed computational cost, whereas for smaller values of $\chi$ the binary scheme is the most cost effective. The crossover occurs at a cost of order $C \approx 10^7$ floating-point operations per iteration, which is well within the capabilities of a small workstation. Therefore, in most practical numerical applications the modified binary MERA is a better scheme than the others.

\begin{table}[!htbp]
\centering
\caption[]{Best fit coefficients to the functional form of Eq. \ref{eq:a2e1} for the scaling of relative energy error in critical ground states, comparing the three different MERA schemes as described in Sect. \ref{sect:ChoiceScheme}. The central charge $c$ of the critical models is given for reference.}
\renewcommand{\arraystretch}{1.2}
\setlength\tabcolsep{5pt}
\begin{tabular}{ |c | r r | r r | r r |}
\hline
$~~~~~$  & \multicolumn{2}{c|}{(i) binary} & \multicolumn{2}{c|}{(ii) ternary} & \multicolumn{2}{c|}{(iii) mod. bin.}  \\
$~~~~~$  & $a~~$ & $b~~$ & $a~~$ & $b~~$ & $a~~$ & $b~~$  \\
\hline
Ising Model ($c=1/2$)            & $~$ 0.012       & 6.20  & $~$ 0.105    & 5.98  & $~~$ 1.13       & 6.81     \\
Potts Model ($c=4/5$)            & $~$ 0.527       & 4.63  & $~$ 1.786    & 4.58  & $~~$ 17.80      & 5.22      \\
XX Model ($c=1$)                 & $~$ 0.615       & 3.97  & $~$ 1.156    & 4.05  & $~~$ 5.25       & 4.30     \\
Heisenberg Zig-Zag ($c=1$)       & $~$ 0.824       & 3.90  & $~$ 3.264    & 3.95  & $~~$ 1.89       & 3.80      \\
\hline
\end{tabular}
\label{tab:SchemeEnergy}       
\end{table}

\section{Reducing the Cost of MERA Contractions} \label{MERAapprox} 

In Appendix \ref{sect:SchemeCompare}, a numerical study determined the modified binary MERA scheme, which can be optimized with a leading computational cost which scales as $O(\chi^7)$, to be the optimal $1D$ MERA implementation of those considered. In this Appendix we describe how the cost of implementing the modified binary MERA scheme can be reduced from $O(\chi^7)$ to $O(\chi^6)$ through the use of approximations in the tensor network contractions required for its optimization. The steps required to optimize the modified binary MERA are analogous to those described for the ternary MERA in Sect. \ref{sect:ScaleImp}. Fig. \ref{fig:ModBinScheme} shows four closed tensor networks (tensor networks without any open index). The tensor networks required to optimize the modified binary MERA (which include the ascending and descending superoperators, and the environments for single tensors) can be generated from these closed tensor networks through removal of a single tensor. The cost of contracting the two networks in Figs. \ref{fig:ModBinScheme}(i-ii) scales as $O(\chi^7)$, whereas cost of contracting the two networks in Figs. \ref{fig:ModBinScheme}(iii-iv) scales just as $O(\chi^6)$ contraction cost. If we could reduce the cost of the two first networks also down to $O(\chi^6)$, then that would be the overall leading cost of the modified binary MERA. Let us then see how this can be accomplished.

\subsection{Insertion of Projectors}

\begin{figure}[!htbp]
\begin{center}
\includegraphics[width=12cm]{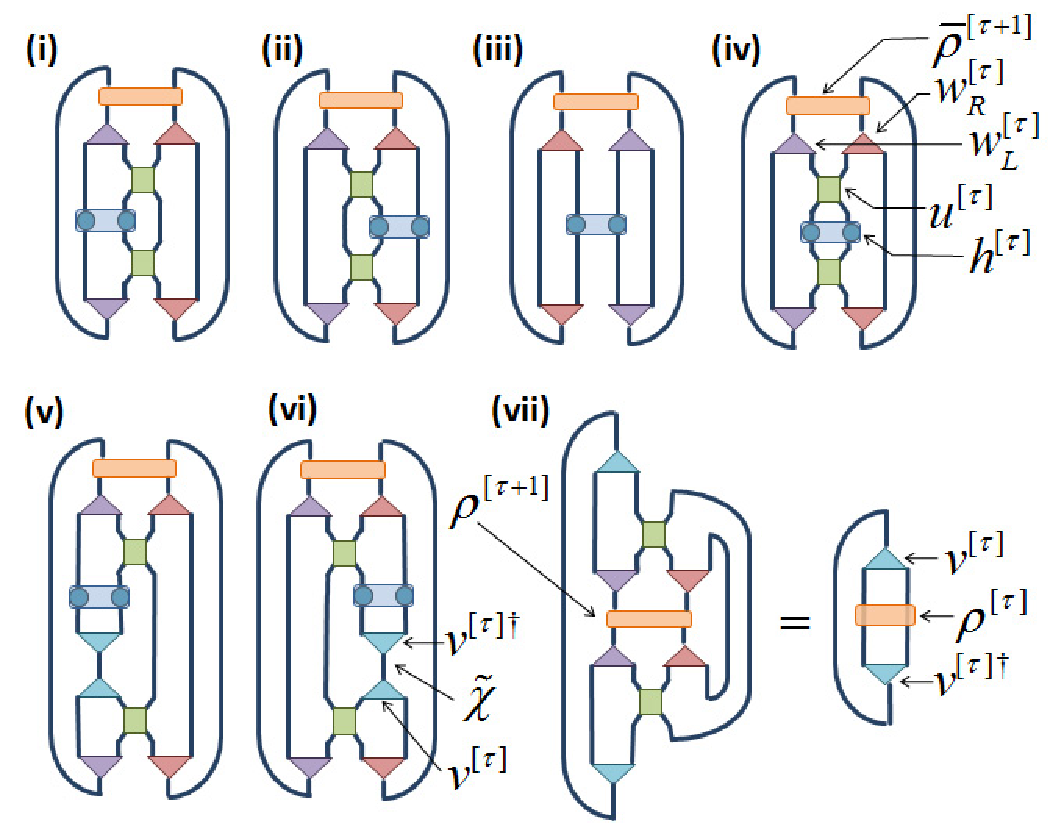}
\caption{(i-iv) The four types of tensor network required for the optimization of the modified binary MERA. In terms of bond dimension $\chi$, the networks of (i) and (ii) are contraction cost $O(\chi^7)$, while networks of (iii) and (iv) are of contraction cost $O(\chi^6)$. (v-vi) The tensor networks of (i) and (ii) have been modified with the inclusion of a rank $\tilde \chi$ projector $P= v v^\dag$. The modified networks are of contraction cost $O(\tilde \chi \chi^5)$. (vii) The tensor $v^{[\tau]}$ should be optimized such that $P^{[\tau]}\equiv v^{[\tau]} v^{[\tau ] \dag}$ projects onto the subspace of the density matrix $\rho^{[\tau]}$ with greatest weight, as described Eq. \ref{eq:a3e2}}.
\label{fig:ModBinScheme}
\end{center}
\end{figure}

Let us consider a rank $\tilde \chi$ projector $P$, decomposed as the product of an isometric tensor $v$ and its conjugate $v^\dag$,
\begin{equation}
P \equiv v v^\dag, \;\;\;\; v^{\dagger}:\left( \mathbb V_\chi \right) ^{ \otimes 2}  \mapsto \mathbb V_{\tilde \chi}, \label{eq:a3e1}
\end{equation}
where $\mathbb V_\chi$ is a $\chi$-dimensional vector space and $\mathbb V_{\tilde \chi}$ is a $\tilde \chi$-dimensional vector space for some $\tilde \chi \le \chi^2$. In place of using the original tensor networks of Fig. \ref{fig:ModBinScheme}(i-ii) in the optimization algorithm, with contraction cost $O(\chi^7)$, we shall use the tensor networks of Fig. \ref{fig:ModBinScheme}(v-vi), which have been modified through inclusion of the projector $P$ and whose cost is now of order $O(\chi^5 \tilde \chi)$. In the reminder of this Appendix we first explain how to optimally choose the isometric $v$ so that the modified tensors networks yield environments that best approximate the environments obtained from the original networks; and then we argue, based on numerical evidence, that $\tilde \chi$ may indeed be chosen as $O(\chi)$ without significant loss of accuracy, thus reducing the overall cost of optimization down to $O(\chi^6)$.

\subsection{Optimizing the Projector} \label{sect:OptProj}

\begin{figure}[!tbp]
\begin{center}
\includegraphics[width=8cm]{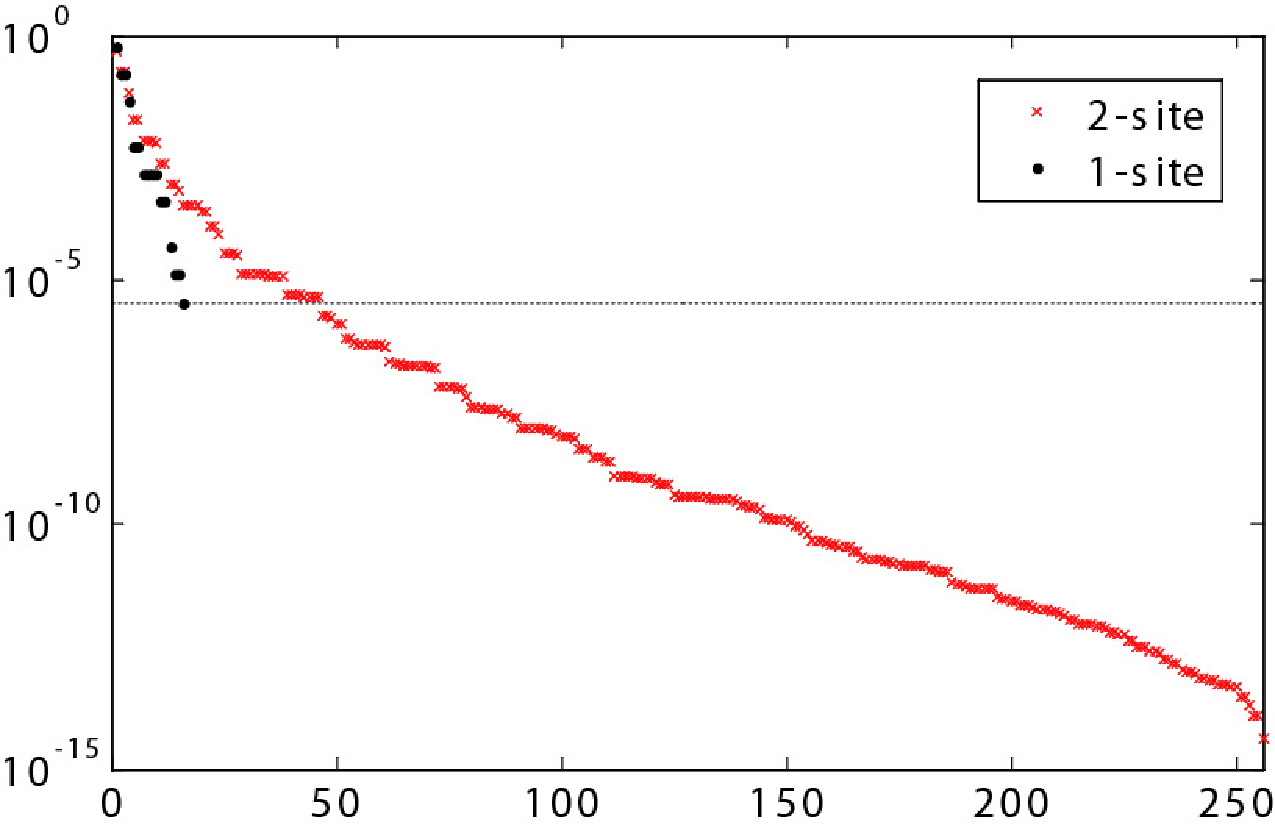}
\caption{The spectrums of one-site and two-site density matrices, obtained from the scale-invariant fixed point of a $\chi=16$ MERA optimized for the ground state of the quantum XX model. The two site density matrix can be truncated down to its $\tilde \chi =50$ most significant eigenvalues, out of the total $\chi^2=256$ eigenvalues, without significant loss of accuracy (as gauged by the smallest eigenvalue of the one-site density matrix).}
\label{fig:DensitySpect}
\end{center}
\end{figure}

Given the two-body density matrix $\rho^{[\tau]}$ at level $\mathcal L^{[\tau]}$ of the MERA, the proper choice for the isometric $v^{[\tau]}$ is that which maximizes the trace of the density matrix,
\begin{equation}
\mathop {\max }\limits_{v^{[\tau ]}} \left( {\rm{tr}}\left( {v^{[\tau]\dag}  \rho ^{[\tau ]} v^{[\tau ]}} \right) \right), \label{eq:a3e2}
\end{equation}
as we now justify. Under this choice of $v^{[\tau]}$ the projector $P^{[\tau]}\equiv v^{[\tau]} v^{[\tau]\dag}$ projects onto the subspace of $\rho^{[\tau]}$ spanned by its eigenvectors of largest eigenvalue. If we then assume that only $\tilde \chi$ eigenvalues out of the total $\chi^2$ eigenvalues of $\rho^{[\tau]}$ have significant weight (the rest being negligibly small) then the density matrix is invariant under the projection, $\rho^{[\tau]}=P^{[\tau]} \rho^{[\tau]}$, to within, by assumption, negligible error. Thus, in this case, when computing the density matrix $\rho ^{[\tau ]}$ from the higher level density matrix $\rho ^{[\tau +1]}$, the modified tensor networks of Fig. \ref{fig:ModBinScheme}(v-vi) will give the same result as the original tensor networks of Fig. \ref{fig:ModBinScheme}(i-ii). Likewise the other tensor environments (such as environments of disentanglers and isometries) generated from the modified tensor networks will be the same as those generated from the original networks.

In principle the isometric $v^{[\tau]}$ that projects onto the most significant subspace of $\rho^{[\tau]}$, as described Eq. \ref{eq:a3e2}, could be obtained directly from the spectral decomposition of the density matrix $\rho^{[\tau]}$ for each $\tau$; however, computing the exact reduced density matrix $\rho^{[\tau]}$ from $\rho^{[\tau+1]}$ is an $O(\chi^7)$ operation and should be avoided. By expressing the density matrix $\rho^{[\tau]}$ in terms of $\rho^{[\tau+1]}$, $u^{[\tau]}$ and $w^{[\tau]}$, we can write the trace in Eq. \ref{eq:a3e2} as the tensor network in the left hand side of Fig. \ref{fig:ModBinScheme}(vii). From that tensor network, we can extract the environment for $v$ with cost $O(\chi^5 \tilde{\chi})$. Therefore, by using the linearized single tensor optimization described in Sect. \ref{sect:LinOpt} we can iteratively optimize $v$ at a cost $O(\chi^5 \tilde{\chi})$.

\subsection{Rank of Projector} \label{sect:RankProj}

The introduction of a properly optimized rank $\tilde \chi$ projector $P$ was argued to be equivalent to truncation of the two-site density matrix to retain its $\tilde \chi$ most significant eigenvalues; this leads to truncation error $\varepsilon  = 1 - {\rm{tr}}\left( {P^{[\tau ]} \rho ^{[\tau ]} } \right)$ that should be kept sufficiently small. A good indication of exactly how small $\varepsilon$ should be kept is to ensure that it is the same size, or smaller than, the smallest eigenvalue of the one-site density matrix, which is indicative of the degree of accuracy of the MERA under consideration. Thus the relevant question becomes, for a MERA with bond dimension $\chi$, what is the necessary rank $\tilde \chi$ of the projector $P$ required to maintain this sufficient degree of accuracy? 

Since the MERA represents states with at most a logarithmic scaling of the entropy with block size, see Sect. \ref{sect:BasicProp}, the entropy of the two-site density matrix is much less than twice the entropy of the one-site density matrix. Therefore, it is to be expected that $\tilde \chi \ll \chi^2$. [Notice that $\tilde \chi = \chi^2$ would be consistent with an entropy that scales linearly with the block size]. Indeed, the numerical evidence suggests that $\tilde\chi$ may be chosen as $O(\chi)$; for the simulations with the modified binary MERA scheme of Sect. \ref{sect:MERABench}, it was found that keeping $\tilde \chi \approx 3 \chi$ gave a sufficient level of accuracy for all four critical models under consideration, and over the large range of bond dimensions $\chi$ analyzed. Given that this relation held over a range of bond dimensions between $\chi=4$ and $\chi=150$, it seems likely that the relation between $\chi$ and $\tilde \chi$ is indeed linear, or at least very close to linear for ground states of critical systems. An example is shown in Fig. \ref{fig:DensitySpect}, which plots the spectrum of one-site and two-site density matrices for $\chi=16$ quantum XX model, where it can seen that choosing $\tilde \chi \approx 3 \chi$ is sufficient to ensure that the truncation error $\varepsilon$ is of the same order as the smallest eigenvalue of the one-site density matrix.

Thus, though not rigorously justified, the available evidence indicates that $\tilde\chi$ may be chosen as $O(\chi)$, hence the overall cost of optimizing the modified binary MERA has been reduced to $O(\chi^6)$ cost. The reduction in cost comes at the price of both introducing new tensors $v^{[\tau]}$ that must be updated with each iteration and introducing a controlled amount of approximation into the tensor network contractions. Although we have only described how the cost of the modified binary MERA scheme can be reduced through introduction of a projector $P$ into the tensor network diagrams, the same approach could be employed to potentially reduce the cost of any MERA scheme.

\end{document}